%% file: dual_CS15.tex
\documentclass[aps,amssymb,amsmath,pra,twocolumn,showpacs,superscriptaddress,longbibliography]{revtex4-2}

\usepackage{graphicx}
\usepackage{dcolumn}
\usepackage{bm}
\usepackage{subfigure}
\usepackage{color}
\usepackage{dsfont}
\usepackage[utf8x]{inputenc} 

\usepackage{amssymb,amsmath,amsfonts,latexsym,graphicx,verbatim}

\usepackage[margin=0.75in]{geometry}

\usepackage[english]{babel}
\usepackage{times}
\usepackage{latexsym}
\usepackage{fancyhdr}
\usepackage{float}
\usepackage{afterpage}
\usepackage{enumitem}
\usepackage{eso-pic, graphicx}

\usepackage{listings}
\usepackage{multirow}

\usepackage{bbm}
\usepackage{upgreek}
\usepackage{amsmath}
\usepackage[T1]{fontenc}

\definecolor{Ablue}{rgb}{0.96,0.24,0.00}

\definecolor{Abluetitle}{rgb}{0.,0.24,0.51}

\definecolor{orange}{rgb}{0.96,0.24,0.00}

\definecolor{darkred}{rgb}{0.55, 0.0, 0.0}
\addto\captionsenglish{\renewcommand{\figurename}{Fig.}}

\definecolor{Gray}{gray}{0.85}
\definecolor{LightCyan}{rgb}{0.88,1,1}
\definecolor{darksalmon}{rgb}{0.91, 0.59, 0.48}
\definecolor{maroon}{cmyk}{0,0.87,0.68,0.32}

\definecolor{mustard}{rgb}{1.0, 0.86, 0.35}
\newcolumntype{a}{>{\columncolor{Gray}}c}
\newcolumntype{b}{>{\columncolor{white}}c}

\usepackage{array}
\newcolumntype{L}[1]{>{\raggedright\let\newline\\\arraybackslash\hspace{0pt}}m{#1}}
\newcolumntype{C}[1]{>{\centering\let\newline\\\arraybackslash\hspace{0pt}}m{#1}}
\newcolumntype{R}[1]{>{\raggedleft\let\newline\\\arraybackslash\hspace{0pt}}m{#1}}

\usepackage[colorlinks=true , citecolor=blue,urlcolor=blue]{hyperref}

\input{Commands3.tex}

\makeatletter
\newcommand*{\rom}[1]{\expandafter\@slowromancap\romannumeral #1@}
\makeatother
\newcommand{\affA}{Department of Chemistry, University of California, Berkeley, California 94720, USA.}
\newcommand{\affB}{Department of Physics, California Institute of Technology, Pasadena, CA 91125, USA.}
\newcommand{\affC}{Chemical Sciences Division,  Lawrence Berkeley National Laboratory,  Berkeley, CA 94720, USA.}

\begin{document}
\title{Dual-space Compressed Sensing}

   \author{Xudong  Lv}\affiliation{\affA}\affiliation{\affB}
	\author{Ashok Ajoy}\affiliation{\affA}\affiliation{\affC}
	
\begin{abstract}
Compressed sensing (CS) is a powerful method routinely employed to accelerate image acquisition. It is particularly suited to situations when the image under consideration is sparse but can be sampled in a basis where it is non-sparse. Here we propose an alternate CS regime in situations where the image can be sampled in two incoherent spaces simultaneously, with a special focus on image sampling in Fourier reciprocal spaces (e.g. real-space and k-space). Information is fed-forward from one space to the other, allowing new opportunities to efficiently solve the optimization problem at the heart of CS image reconstruction. We show that considerable gains in imaging acceleration are then possible over conventional CS.  The technique provides enhanced robustness to noise, and is well suited to edge-detection problems. We envision applications for imaging collections of nanodiamond (ND) particles targeting specific regions in a volume of interest, exploiting the ability of lattice defects (NV centers) to allow ND particles to be imaged in reciprocal spaces simultaneously via optical fluorescence and $\Cs$ magnetic resonance imaging (MRI) respectively. Broadly this work suggests the potential to interface CS principles with hybrid sampling strategies to yield speedup in signal acquisition in many practical settings. 
\end{abstract}

\maketitle

\begin{figure}[t]
  \centering
  {\includegraphics[width=0.45\textwidth]{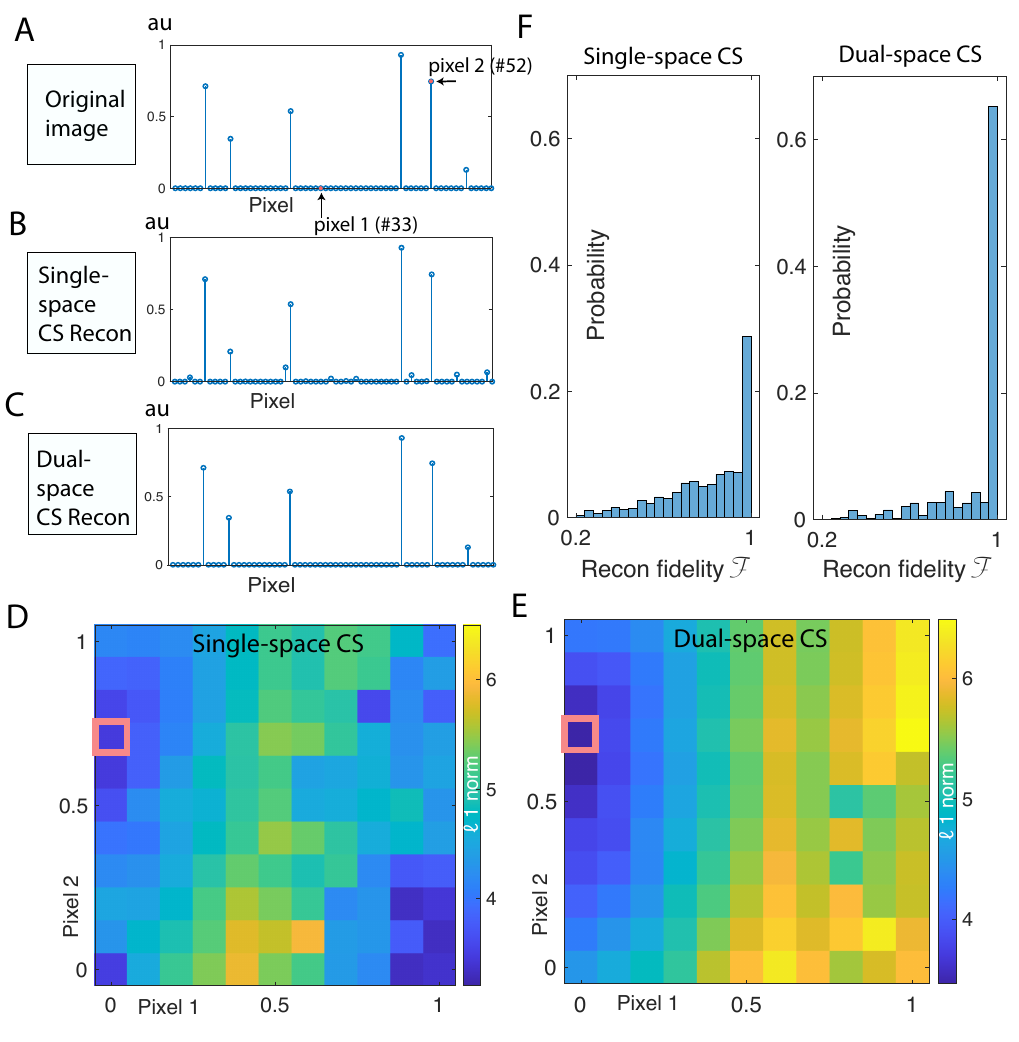}}
  \caption{\T{Reconstruction with single-space and dual-space CS.}
(A) \textit{Illustrative 1D image} consisting of $\xd$-functions with $n{=}64$ pixels and sparsity $s{=}6$. (B) \textit{Conventional CS reconstruction} employing $m{=}$9 random samples in k-space fails to  converge to the original image. (C) \textit{Dual-space CS}, carried out with the same number of measurements $m_{\text{dual}}{=}m_k{+}m_x{=}m$,  leads to closer reconstruction.  (D-E) \textit{Representative landscapes of the optimization parameter space} for single-space and dual-space CS respectively. Axes represent the normalized values for two chosen pixels (marked in (A)), and color denotes corresponding $\| \mathcal{I}\|_{1}$. Red box indicates the true value corresponding to (A). Search landscape of single-space (conventional) CS yields multiple local minima. In contrast, dual-CS has a better conditioned search landscape with a single global minimum. (F) \textit{Distribution of reconstruction fidelity} $\mathcal{F} {=} \braket{\mI}{\bar\mI}$, via statistics performed on 1000 random manifestations of images  as in (A). Dual-space CS shows considerably better reconstruction fidelity.}
\zfl{1d_support}
\end{figure}

\T{\I{Introduction}} --
 Compressed sensing (CS) \cite{donoho2006compressed,candes2006robust,Donoho09, ganguli2010statistical, baraniuk2007compressive, andrade2012application} is used on a daily basis to provide data acquisition speedup in various experimental scenarios spanning imaging \cite{cleary2021compressed,ma2021high, zhu2012faster}, spectroscopy \cite{lu2019compressed} and quantum metrology \cite{arai2015fourier}. CS is particularly effective in settings where the image (data) $\mI$ is sparse in a desired basis $\Phi$, but can be \I{sampled} in an alternate basis $\ov{\Phi}$, that is incoherent to the first, and where the image is non-sparse~\cite{donoho2001uncertainty,Foucart13}. We refer to sparsity $s$ here as the number of non-zero entries in $\mI$, and higher sparsity corresponds to when the image is composed of a large number of zeros. Considering an $n$-dimensional image $\mI$ represented as a column vector, if $b$ is a vector of $m$ random measurements, with $m{<}n$, and $\T{A}$ is a ($m{\zt} n$ dimensional) matrix, CS entails solution of the basic pursuit (BP) problem~\cite{baraniuk2007compressive}, 
\beq
\min _{\mathcal{I}}\left\{\| \mathcal{I}\|_{1}: \T{A} \mathcal{I}= b\right\}\:.
\zl{LP}
\eeq
Here the minimization is carried out over the $\ell_1$-norm, where  $\|\mI\|_1{=}\sum |\mI_j|$, is the sum of the absolute values of pixels in $\mI$. \zr{LP} has been shown to a good proxy to the more accurate, albeit intractable, $\ell_0$ minimization problem~\cite{candes2006robust}. As elucidated by Tao and co-workers~\cite{candes2006robust}, it can be solved efficiently under conditions of high sparsity (small $s$). The minimum number of samples,  $m^*$, required for a high-fidelity reconstruction scales as $m^*{=}Cs\log(n)$~\cite{candes2006robust}, where $C$ is a constant. This reduction of sampling requirements from the dimension of the entire space $n{\rt} \log(n)$ is where the power of CS resides; the reconstruction error scales as $\mO(n^{- g(C)})$, where $g$ is a function~\cite{candes2006robust}.

An important real-world application of CS is in magnetic resonance imaging (MRI~\cite{Liang00}), where it is routinely employed to accelerate image acquisition~\cite{lustig2007sparse}. Image sampling here occurs in k-space; this reciprocal space $\ov{\Phi}$ is {maximally} incoherent to the original real-space image basis~\cite{SOM}. In \zr{LP} then,  $\T{A} {=} \T{R}{\cdot} \T{\R{DFT}}$, where $\T{\R{DFT}}_{jk} {=} (1/\sq{n})e^{-2 \pi ijk / n}$ the $(n{\zt}n)$ Discrete Fourier Transform matrix, and $\T{R}$ is a $(m{\zt}n)$  selection matrix corresponding to the random measurements $b$. Dramatic acceleration factors  ${>}10^4$ have been demonstrated for MRI in sparse image settings~\cite{lustig2008compressed,lustig2007sparse}.

All said however, CS still implicitly assumes that image sampling occurs in only \I{one} space, $\ov{\Phi}$. In this paper, we consider an alternate scenario, assuming the regime where the object can be sampled in \I{both} $\Phi$ and $\ov{\Phi}$ simultaneously. We show that, in this case, considerable further gains are possible compared to conventional CS,  the required number of samples scaling as $m^*{\app}Cs_T\log(n)$, where the sparsity of the new problem,$s_T{<}s$. Since this CS gains obtained from sampling in two spaces, we will refer to it as \I{``dual-CS"}.

The feasibility of dual-space sampling has been recently demonstrated for certain nanoparticle materials that permit their imaging in two Fourier reciprocal spaces \cite{lv2021background}. We are interested here in the rapid reconstruction of images consisting of distributions of such nanoparticles. An example is that of targetable diamond nanoparticles with Nitrogen Vacancy (NV) defect centers \cite{Maze08, miller2020spin}. A fortuitous combination of spin and optical properties \I{(i)} endow NV centers with optical fluorescence that is bright and non-blinking, and \I{(ii)} permit the NV centers to spin polarize surrounding $\Cs$ nuclei in the lattice \cite{fischer2013bulk,ajoy2018ori, Ajoy18, schwartz2018robust,ajoy2020room}. The latter permits the particles to be imaged efficiently via $\Cs$-MRI~\cite{lv2021background}. Such dual-mode optical (real-space) and $\Cs$-MRI (k-space) imaging forms the basis for protocols we consider in this paper.

We show that, in the case,  \zr{LP} is modified as $\min_{\mI}\{\|\mI\|_1{:} \T{A}_k\mI{=}b_k, \T{A}_x\mI{=}b_x\}$, where $b_x$ and $b_k$ refer to measurements in real and k-space respectively. With an appropriately chosen sampling program, this can be recast to
\beq
\min_{\mI_T}\{\|\mI_T\|_1{:} \T{A}_T\mI_T= b_T\},
\zl{truncated_CS}
\eeq 
where $\mI_T$ is a \I{truncated} image with an effectively \I{enhanced} sparsity, $s_T{<}s$. Convergence to the true image is then significantly accelerated (see \zfr{1d_support}) -- this is the central result of this paper.

\begin{figure}[t]
  \centering
  {\includegraphics[width=0.5\textwidth]{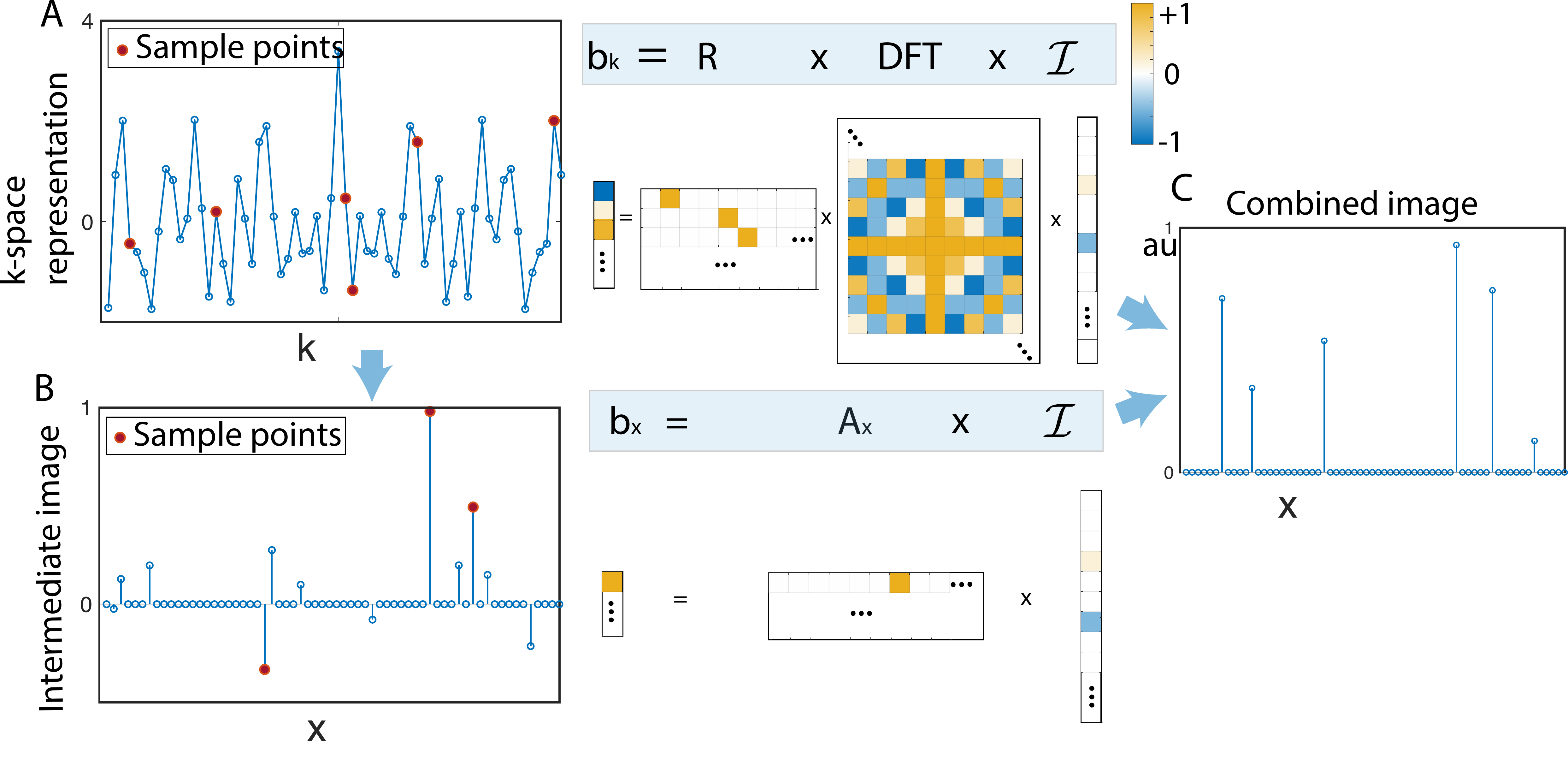}}
  \caption{\T{{Dual-space CS program.}} Steps involved in dual-CS reconstruction for the example image in \zfr{1d_support}A. (A) \T{I}: \textit{k-space sampling.} First, $m_k{=}6$ points are randomly sampled in k-space (blue line) of the $n{=}64$ pixel image $\mathcal{I}$. Right panel: matrix form of the right of \zr{LP}. Color represents a non-zero value of the matrix element (see SI~\cite{SOM}). (B) \T{II}: \textit{Reconstruction and real-space sampling.} An intermediate image (blue line) is reconstructed using data acquired in (A) and solving \zr{LP}. \T{III}: Now, real-space sampling is performed on this intermediate image; $m_x{=}3$ samples here yield $t{=}2$ non-zero coefficients (cf. \zfr{1d_support}A), which is extracted from the image, yielding $\mI_T$ with enhanced sparsity. Right panel: analogous matrix representation for x-space sampling. (C) \T{IV:} \textit{Final reconstructed image} obtained via \zr{truncated_CS} for $\mI_T$ yields the dual-CS reconstructed image $\bar \mI $.} 
\zfl{protocol}
\end{figure}

\T{\I{Principle}} -- 
The origin of the speedup and exact sampling algorithm is best explained with an illustrative example, shown in \zfr{1d_support}. We consider a simple 1D image $\mI$ (\zfr{1d_support}A) consisting of a collection of sparse objects ($\xd$-functions), e.g. single ND particles. Conventional CS via k-space sampling leads to the reconstruction as shown in \zfr{1d_support}B, where we solve the BP problem in \zr{LP} through linear programming (LP) by defining a set of parameters $\mathbf{u}$ and $\mathbf{v}$~\cite{eldar2012compressed}:
\beq
u_{i}=\max \left\{\mathcal{I}_{i}, 0\right\}, v_{i}=\max \left\{-\mathcal{I}_{i}, 0\right\}, \mathcal{I}=\T{u}-\T{v}
\zl{LP3}
\eeq
reformulating the problem as $\min _{u, v} \sum_{i=1}^{n} (u_{i}{+}v_{i})$,
subject to
$A(\mathbf{u}-\mathbf{v}){=}b\:;\:\mathbf{u} {\geq} 0\:;\:\mathbf{v} {\geq} 0$. While close, the reconstructed image $\ov{\mI}$ deviates from the true image on account of the BP problem failing to converge to the true global minimum (\zfr{1d_support}B).  Instead dual-CS (\zfr{1d_support}C), with same number of measurements, as we shall describe shortly, can yield much closer convergence.

Before a detailed discussion, and to intuitively motivate the basis for this improvement, let us first consider the landscape of the BP optimization process during conventional (singel-space) CS reconstruction (\zfr{1d_support}B). For simplicity, \zfr{1d_support}D considers a pair of pixels in image $\mI$ (marked in \zfr{1d_support}A). For a (normalized) numerical choice of these pixel values from 0 to 1 (shown as the axes), the colors represent the corresponding $\ell_1$-norm $\| \mathcal{I}\|_{1}$. \zfr{1d_support}D therefore reflects the landscape of the optimization process, but focused here only on the effect of a pair of pixels for convenience; in general the landscape occupies a multidimensional ($\mathbb{R}^{n}$) space.  Already, however, \zfr{1d_support}D illustrates that conventional-CS optimization landscape has several local minima. In comparison however, the landscape is much better conditioned in the dual-CS case (\zfr{1d_support}E), and prompts more accurate minimization in the latter case.  \zfr{1d_support}F illustrates this from a complementary viewpoint. For 1000 random manifestations of the image $\mI$ as in \zfr{1d_support}A, we plot a histogram of the reconstruction fidelity $\mathcal{F}{=}\braket{\ov{\mI}}{\mI}$, showing significantly lower reconstruction error in dual-CS. 

Let us now elucidate the exact sampling \I{program} (algorithm) employed for dual-CS (\zfr{protocol}). We use, for illustration, an image consisting of $\delta$-function objects (peaks), as in \zfr{1d_support}A: (\T{\rom{1}}) First, we perform a k-space CS reconstruction using $m_k$ samples (\zfr{protocol}A). (\T{\rom{2}}) We then select a few ($m_x$) points from the intermediate image, and sample them \I{additionally} in real-space (\zfr{protocol}B). While in principle, any choice of the reconstructed peaks is suitable, we focus on the $m_x$ peaks with largest values identified in the intermediate image.  (\T{\rom{3}}) Upon real-space sampling, a few pixels (peaks) are fully determined in the image. We can then remove them to form a truncated image $\mI_T$ with sparsity $s_T$. Note however, that real-space sampling might not yield a valid image peak for each of the $m_x$ measurements; we define a corresponding \I{success probability} as $\alpha_x{=}\frac{s-s_T}{m_x}$, where the right hand side refers to the number of true image peaks identified per x-space measurement. (\T{\rom{4}}) Finally, the LP problem is re-solved again for  $\mI_T$ following \zr{LP3} to yield the reconstructed image $\bar \mI$ (\zfr{protocol}C).

This program yields an enhanced rate of convergence (\zfr{1d_support}F). Intuitively, the origin of the resulting speedup is simple to elucidate. Real-space sampling permits the ability to localize a few of the true image pixels rapidly and with a high degree of confidence. Effectively, the image is decomposed into, $\mI{=} \mI_T{\oplus} \mI_x$, where $\mI_x$ is obtained from real-space sampling, and the sparsity in $\mI_T$ is enhanced. Compared to conventional CS, real-space sampling provides specific information of a few image pixels and reduces the size of the search space required for image reconstruction. 

We can quantify conditions under which the dual-CS can provide an advantage. 
Consider that reconstructing the image with the above program requires a total number of measurements that is at least,
\beq
m^*_{\text{dual}} = C(s-\alpha_x m_x)\log n + m_x \:.
\zl{dual_measure}
\eeq
In contrast, conventional k-space CS requires $m^* {=} Cs\log n$ measurements. Dual-CS requires fewer measurements, and hence provides a speedup, if $m^*_{\text{dual}} {<} m^*$, or equivalently, $C(s-s_T)\log n>m_x$, or,  $\frac{s-s_T}{m_x}>\frac{s}{m^*}$.  The left hand side here is the x-space success probability $\alpha_x$. In an analogous manner,  we identify the right hand side $\alpha_k {=} \frac{s}{m^*}$ as the k-space success probability -- effectively the information obtained by sampling one k-space point. Hence dual-space CS requires less measurements to achieve the same reconstruction probability when  $\alpha_x {>} \alpha_k$, or when the success rate of x-space sampling is higher than k-space. Intuitively, this condition would be satisfied when the prior knowledge extracted from the step (\T{\rom{1}}) of the program facilitates well-focused x-space sampling in step (\T{\rom{2}}). In contrast, however, in single-space CS, k-space samples are chosen randomly by definition. Given the different nature of sampling -- targeted versus random -- there is a wide portion of the landscape where this condition is satisfied.

\begin{figure}[t]
  \centering
  {\includegraphics[width=0.48\textwidth]{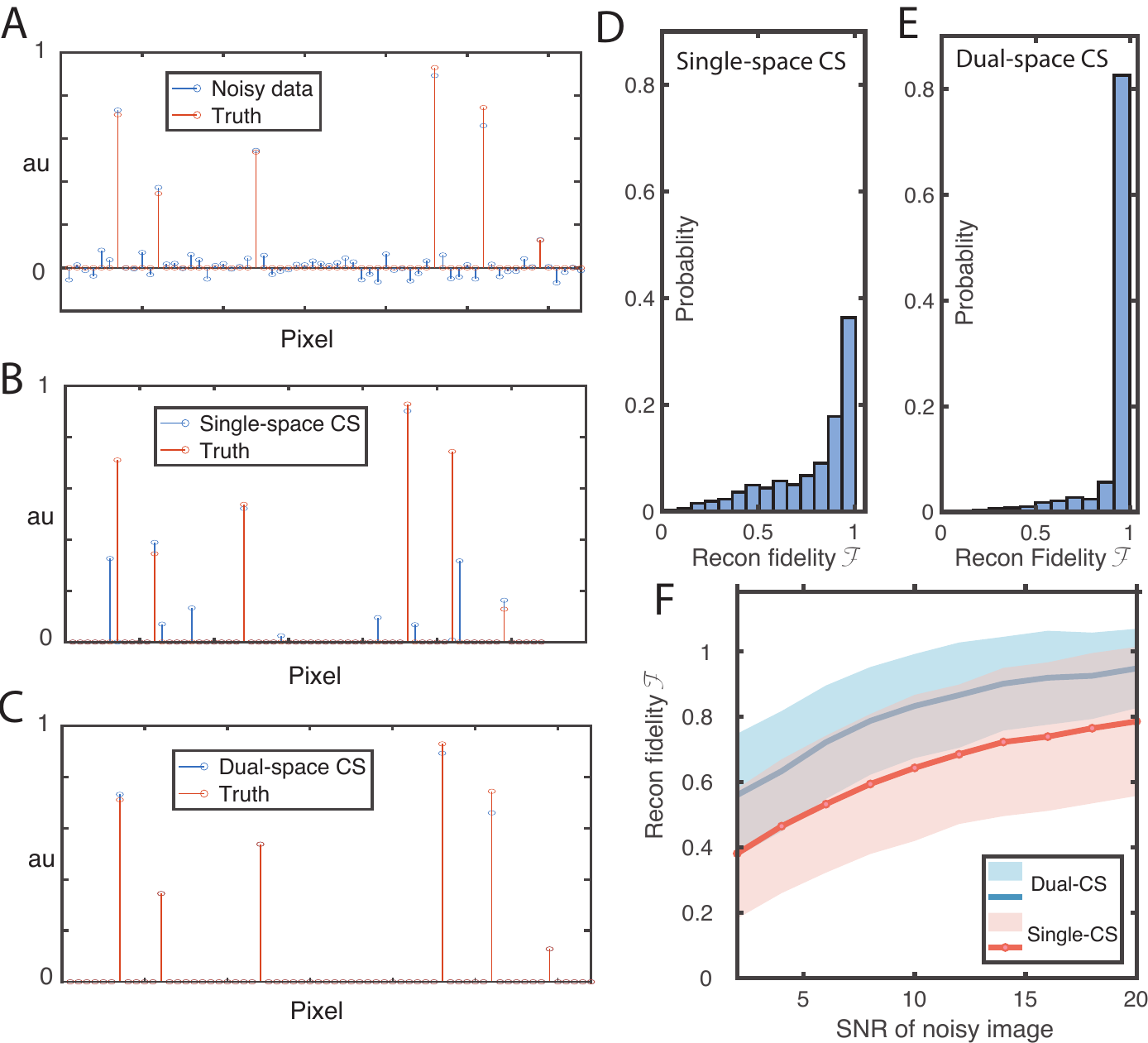}}
  \caption{\T{{Application to noisy image reconstruction.}} (A) Representative image from \zfr{1d_support}A (red) deteriorated by addition of random Gaussian noise. (B) \textit{Single-space CS reconstruction} of the  noisy image in (A) yields deviations from the true image $\mI$. (C) \textit{Dual-space CS reconstruction}, in contrast, shows improved reconstruction fidelity. (D-E) \textit{Reconstruction fidelity  $\mF$ distribution} for single- and dual-space CS cases obtained by random sampling of noisy images (with SNR${=}20$) (cf. \zfr{1d_support}F). (F) \textit{Reconstruction fidelity plotted against image SNR} shows dual-CS consistently performs better than conventional CS. Shaded region represents standard deviation obtained from statistics over 1000 random manifestations of images as in \zfr{1d_support}A.}
\zfl{Noise_recon}
\end{figure}

\T{\I{Robustness in noisy settings}} -- 
Apart from higher reconstruction fidelity, dual-CS is also effective in realistic scenarios when the image of interest is admixed with noise. \zfr{Noise_recon}A shows one such example, where we consider the image with $\delta$-function peaks in \zfr{1d_support}A deteriorated by Gaussian white noise (with variance $\sigma^2$). The noisy image is denoted as $\mI'$, and we assume as in \zfr{Noise_recon}A that some of the peak pixels are higher than the noise level, i.e. have an signal to noise ratio (SNR) greater than 1. In conventional CS, for a randomly selected measurement, the effective SNR is ${\approx} \frac{1}{n}\frac{\sum |\mI_i|}{\sigma}$, where $i$ represents the pixel index. In contrast, in step \T{(\rom{2})} of the program above, we sample the intermediate image in real-space, allowing a better localization of image peaks. In this case, the intermediate image SNR is ${\approx} \frac{1}{m_x}\frac{\sum_{U} |\mI_i|}{\sigma}$, where $U$ represents a set of $m_x$ pixels, and $\sum_{U} |\mI_i|{\approx} \frac{s{-s_T}}{s}\sum |\mI_i|$ for a random manifestation of the image. Effectively therefore, the SNR is ${\approx} \frac{\alpha_x}{s}\frac{\sum |\mI_i|}{\sigma}$ due to the x-space sampling in step \T{(\rom{2})}. Since $\frac{\alpha_x}{s} {>} \frac{1}{n}$ (given $\alpha_x {>} \frac{s}{m^*} {>} \frac{s}{n}$), there is a boost in SNR from the guided x-space sampling. Intuitively, this gain arises because a larger portion of the samples here are in close proximity to the true image peaks. 

\zfr{Noise_recon}B-C demonstrates this by comparing image reconstruction under conventional CS and dual-CS in noisy settings. In the former, noise leads to spurious peaks in the reconstructed image (\zfr{Noise_recon}B).  In contrast, with the addition of real-space sampling in dual-CS, we are able to faithfully locate a subset of the peaks. Eliminating them to form the truncated image $\mI_T$ and re-solving the BP problem then leads to a closer match with the true image (\zfr{Noise_recon}C). The histogram in \zfr{Noise_recon}D shows the image reconstruction fidelity is better with dual-CS compared to single-CS (\zfr{Noise_recon}E). Finally, \zfr{Noise_recon}F shows the reconstruction fidelity with average image SNR. Clearly dual-CS shows enhanced performance,  including down to where the SNR of the image pixels approaches 1.

\T{\I{Application to edge detection}} --
 Dual-CS, as we have described it, can be generalized to serve other ``dual" spaces apart from x- and k-spaces, with the proviso that they are incoherent with respect to each other. As an example, we focus on its application to edge detection \cite{zhou2020flat}, where two spaces are considered \I{finite-difference} (FD) space and k-space. The setting we have in mind is that of regions of disconnected filled volumes with the goal of determining the boundaries between them~\cite{canny1986computational}. Practically this can refer, for instance, to nanodiamond (ND) particles targeting disjoint regions; the ND particles being imaged by optics and $\Cs$ MRI simultaneously.  At the outset, it is clear that conventional CS via k-space sampling is not particularly suited for this problem. Edge information resides in the high k-orders and sampling an adequate amount of points with large k-coefficient requires significantly more samples (scaling as ${\app}\pi k^2$). In conventional CS, subsampled k-space data usually enables reconstruction of low frequency {``bulk"} shape of an object, while lacking the faithful high frequency {``edge"} profile because the algorithm optimizes towards global image fidelity. In contrast, dual-space sampling can provide a significant acceleration, since the bulk regions, reconstructed via a few k-space samples, can be sampled again in real-space to discern the region boundaries. Such secondary sampling, by its very nature, can access high frequency components of an image with only small number of samples assuming the broad layout is pre-determined via original k-space reconstruction.

The edge detection problem can be reformulated and solved with the program described above. Assume the image of interest can be decomposed into spatially slow varying and fast varying components, corresponding to the bulk and edge components of the image. Defining the local gradient $ \nabla \mathcal{I}(x)$, these components are assumed to satisfy $\nabla \mathcal{I} {\leq} G_0$ and $\nabla \mathcal{I} {\geq} G_0$, where $G_0$ is a threshold (see \zfr{demo}). Such images have a convenient sparse representation in the FD space, defined here as being obtained by the transformation $\mathcal{FD}[\mathcal{I}(x)] {=} \mathcal{I}(x{+}\Delta x) {-} \mathcal{I}(x)$ -- the difference between adjacent pixels, where $\Delta x$ is the pixel size. High sparsity in this space arises because isolated regions with relatively uniform intensity correspond to $\mathcal{I}(x{+}\Delta x) {-} \mathcal{I}(x) {\approx} 0$ for most $x$. Edge reconstruction can be carried out by reconstructing the $s$ sparse peaks \I{in FD space} via k-space and real-space measurements. FD space is sampled indirectly by measuring adjacent pixels in x-space and then taking the difference. The program is analogous to the one previously described, with the difference being that step (\T{\rom{4}})  involves solving a modified BP problem, $\min_{\mI}\{\|\mathcal{FD}[\mI]\|_1{:} \T{A}_k\mI{=}b_k, \T{A}_{\text{FD}}\mathcal{FD}[\mI]{=}b_{\text{FD}}\}$, where $\T{A}_{\text{FD}}$ and $b_{\text{FD}}$ refer to sampling matrix and sampled outcome in FD space respectively.

\zfr{demo} illustrates such edge detection; a 256${\times}$256 Shepp-Logan phantom (see \zfr{demo}B, C), with a sparsity $s {=} $1726 (in the FD space) is used as an example. We randomly sample $m_k {=} $ 1311 k-space points. After $m_{\text{FD}} {=} 500$ measurements in FD space based on local gradient, the same truncated optimization using data in two spaces yields an image that highly resembles the original for image fidelity $\mathcal{F}{>}$99.9\% and edge fidelity ${>}$99.5\% (\zfr{demo}B, C). Conventional CS yields much worse reconstruction fidelity in comparison.

\begin{figure}[t]
  \centering
  {\includegraphics[width=0.50\textwidth]{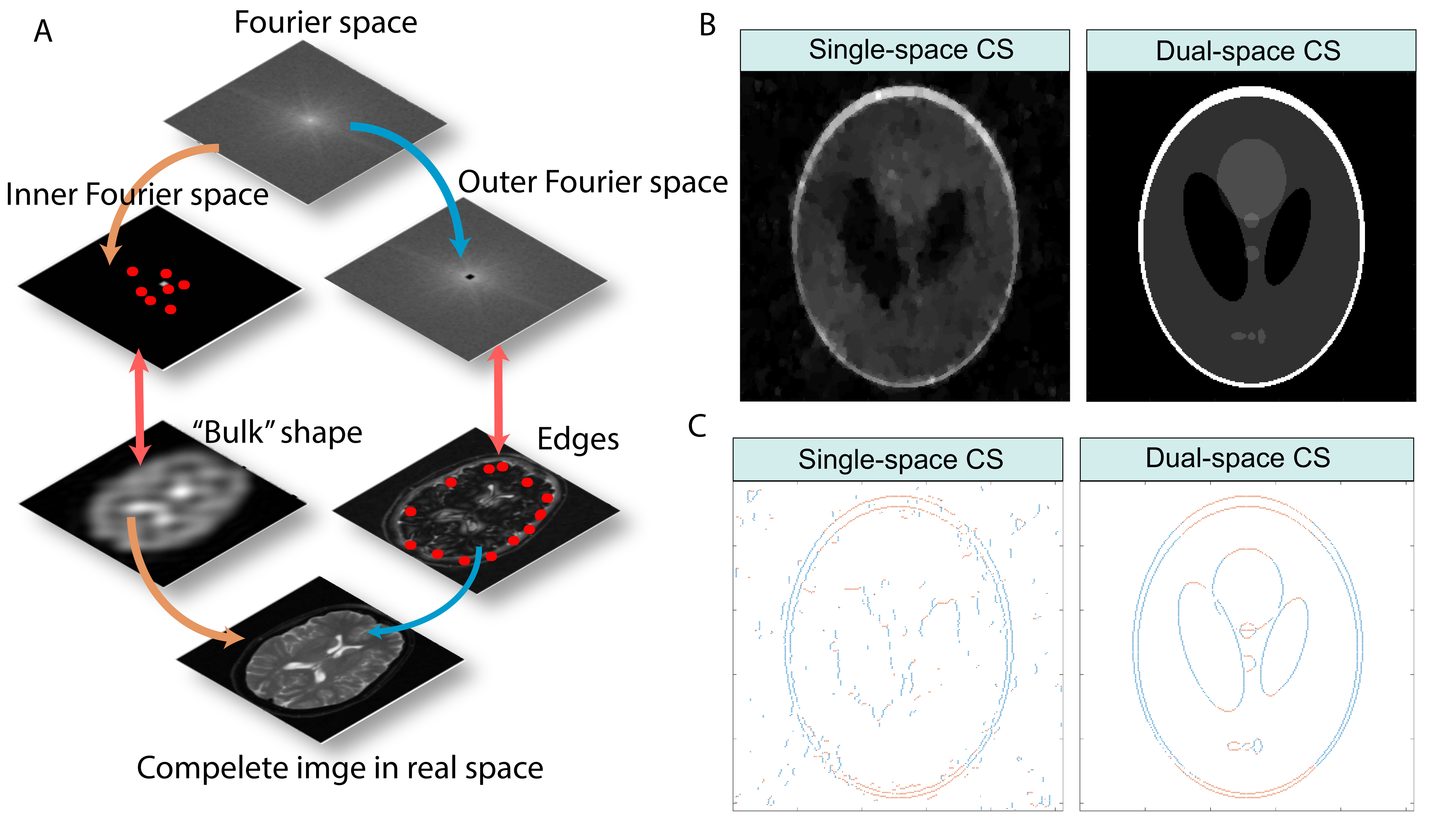}}
  \caption{\T{{Edge detection using two-dimensional dual-space CS}} (A) \I{Strategy.} Fourier-space representation of an image is decomposed into low frequency (inner k-space) and high frequency (outer k-space) components, corresponding to image ``bulk" and ``edges" respectively in real-space. Red points illustrate sampling points in two spaces via the dual-space CS program (\zfr{protocol}). (B) \I{Image reconstruction with single- and dual-space CS.} A 256$\times$256 Shepp-Logan phantom is employed with same number of samples ($m {=} m_x {+} m_k {=} $1811 samples) for both cases. In the dual-space CS, $m_k {=} 1311$ and $m_x {=} 500$. (C) \I{Edge detection with single- and dual-space CS.} Data points corresponding to edge pixels identified by the two approaches respectively (here $G_0{=}0.52$).  Red (blue) color indicates edges with large gradient on vertical (horizontal) direction. Dual-CS consistently identifies edges with higher fidelity with compared to conventional CS. }
\zfl{demo}
\end{figure}

\T{\I{Applications}} -- While the methodology here is developed as a stand-alone theoretical proposal, we anticipate potential applications in two complementary directions. The first involves targeted ND imaging (as described) wherein the ND particles can be imaged through optics and $\Cs$ MRI at low magnetic fields~\cite{lv2021background}. Given typical imaging scenarios operational in in-vitro cellular imaging -- \I{(i)} sparse particles dispersed in a wide field of view, and \I{(ii)} low-power optical imaging through a scanned (rastered) optical beam -- the CS technique developed here might find direct utility.  

A second, complementary, application is in rapidly localizing positions of sparsely distributed NV centers in a nanoscale volume for quantum sensing~\cite{Arai13}. Searching for NV centers in the lattice is a routine but important task; and accelerated localization would allow for a greater number of NV centers that can be screened for useful parameters (e.g. coherence time or surface proximity). However, typical methods of confocal microscopy employed for this task are time-consuming, especially in 3D. Recent work has demonstrated exploiting nanoscale gradients to image the NV centers via a phase-encoding protocol employing conventional CS~\cite{arai2015fourier,Magesan13}.  Our work suggests the possibility of even further acceleration stemming from dual-CS reconstruction.  One can similarly imagine applying this method to imaging Rydberg atoms trapped by an optical tweezer array~\cite{endres2016atom,barredo2018synthetic,dhordjevic2021entanglement} over mm-length scales via a combination of 2f Fourier imaging system followed by sampling with high a NA objective. The imaging speedup might benefit readout fidelity in quantum simulator platforms, given short vacuum lifetimes of the Rydberg atoms. 

We also envisage uses in quantum information (QIS) scenarios, such as entanglement characterization~\cite{howland2013efficient}, simultaneous measurement of complementary observables~\cite{howland2014simultaneous} and state tomography~\cite{huang2020predicting,huang2021provably,vermersch2018unitary}, where recent work has demonstrated gains from exploiting CS for these applications~\cite{wang2018multidimensional,cramer2010efficient,gross2010quantum,shabani2011efficient, kalev2015quantum}. 

\T{\I{Conclusion}} -- In summary, we have proposed and analyzed a strategy for compressed sensing in scenarios where the image can be sampled simultaneously in two incoherent spaces. We focused on the exemplary case of sampling in Fourier reciprocal spaces (real- and k- space), and elucidated how it allows a more efficient solution of the CS reconstruction problem -- yielding enhanced image convergence while also being robust against noise. We envision applications to edge detection, and nanodiamond imaging in real-world media~\cite{lv2021background}. Simultaneously combining information from two incoherent spaces can inspire new measurement approaches for both classical imaging as well as QIS applications.

A.A. acknowledges funding from ONR (N00014-20-1-2806) and DOE STTR (DE-SC0022441).

\bibliography{masterbib}

\clearpage
\onecolumngrid
\begin{center}
\textbf{\large{\textit{Supplementary Information} \\ \smallskip
Dual-space Compressed Sensing}}\\
\hfill \break
\smallskip
Xudong Lv$^{1,2}$ and Ashok Ajoy$^{1,3}$\\
${}^{1}$\I{{\small Department of Chemistry, University of California, Berkeley, Berkeley, CA 94720, USA.}}\\
${}^{2}$\I{{\small Department of Physics, California Institute of Technology, Pasadena, CA 91125, USA.}}\\
${}^{3}$\I{{\small Chemical Sciences Division Lawrence Berkeley National Laboratory,  Berkeley, CA 94720, USA.}}\\
\end{center}

\twocolumngrid
In this supplementary material, we provide additional information for the article \I{``Dual-space Compressed Sensing"}. In Sec. \ref{2CS}, we provide an overview of conventional (single-space) CS, and contrast it with the dual-space CS proposed here. Sec. \ref{intuitive} provides three complementary view-points that highlight an intuitive understanding of the dual-CS methodology and the origin of imaging speedup. Subsequently, Sec.~\ref{sec_nsp} provides a more detailed proof of conditions under which dual-CS shows enhanced performance over single-CS. Applications to edge detection are detailed and discussed in Sec.~\ref{sec_edge}. We finally provide more numerical results that analyze performance in Sec.~\ref{sec_performance}.

\tableofcontents
\section{Two CS approaches}
\label{2CS}
Here we provide an overview of conventional (single-space) CS and highlight the differences in with dual-CS.

\subsection{Conventional CS} 
Let us consider conventional CS applied in the context of MRI to elucidate its principle. CS is able to reconstruct an real-space image $ \mathcal{I}$ with a dimension of $n$ using only $m(<n)$ of randomly measured samples in k-space. Without losing generality, we assume that the image $ \mathcal{I}$ (in real-space) is sparse, namely the number of non-zero pixels (or ``sparsity") $s$ satisfies $\frac{s}{n} \ll 1$. For instance, for the specific example of Fig. 1A considered in main text \zfr{1d_support}A, there are only 6 non-zero pixels in a 64-pixel image. 

The transformation between image $\mathcal{I}$ and the k-space measurements $b$ is $\T{A} \mathcal{I}=b$ (shown in \zfr{CS_equation_fig}), where the \textit{sampling matrix} $\T{A}$ is a $m\times n$ (row-selected) Fourier matrix. More precisely, $\T{A}$ is composed of a product of two matrices -- $\T{R}$, a $m\times n$ random selection matrix, and $\T{DFT}$, a $n\times n$ discrete Fourier transform matrix. For example in \zfr{CS_equation_fig}, if k-space pixels indexed 2, 5, 6, 9 are sampled, one could draw rows with the same indices from an identity matrix to form $R$. Here DFT takes the form of DFT$_{jk} = \left(\frac{\omega^{j k}}{\sqrt{n}}\right)_{j, k=-\frac{-n+1}{2}, \ldots, -\frac{n-1}{2}}$, where $\omega=e^{-2 \pi i / n}$.

\begin{figure}[h]
  \centering
  {\includegraphics[width=0.46\textwidth]{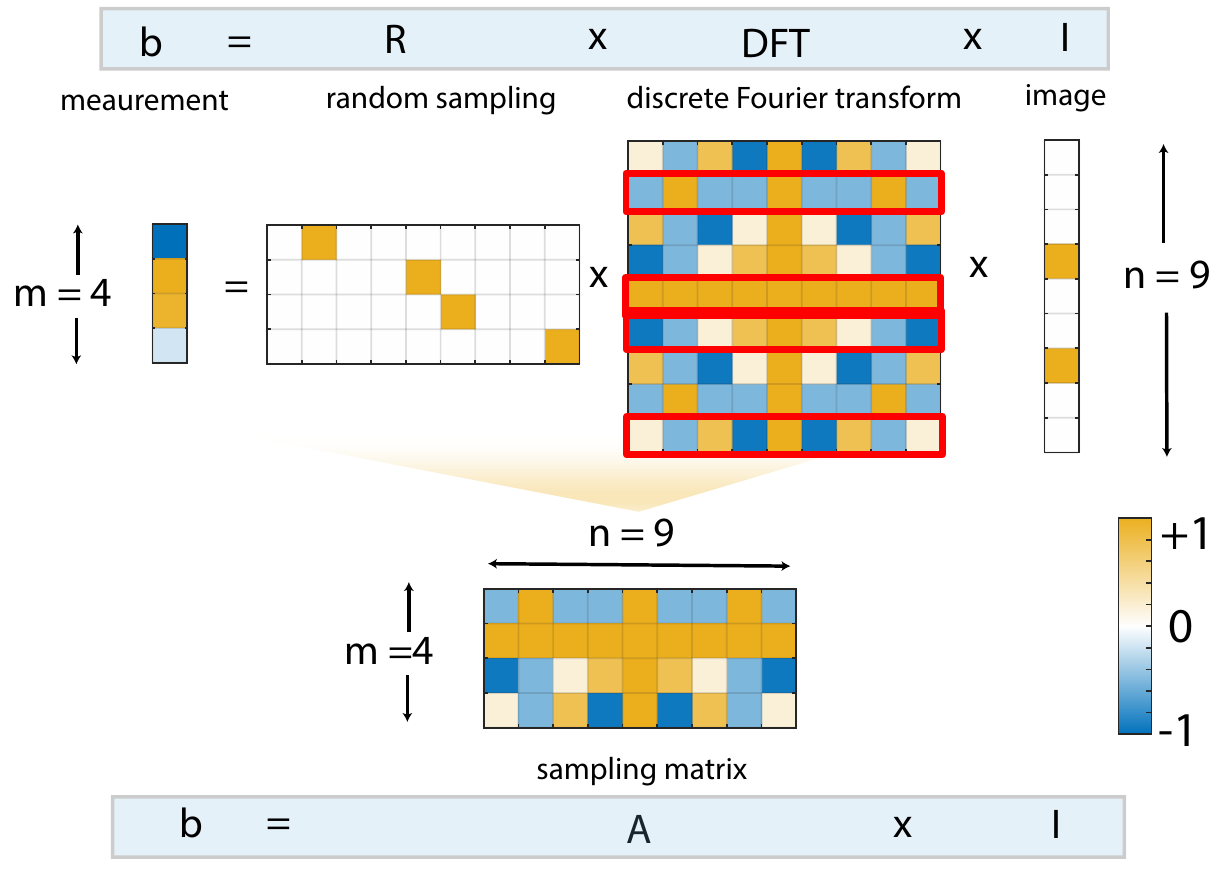}}
  \caption{\T{Schematic illustration of sampling equation} $\T{A} \mathcal{I}=b$ employed for conventional CS reconstruction shown in a matrix format, where the values are represented graphically (see colorbar). In this example, $m=4$ measurements are employed  to acquire signal $b$ from a $n=9$ pixel image $\mathcal{I}$ with assumed sparsity $s=2$. \textit{Top panels:} Sampling process represented by matrix $\T{R}$ randomly picks $m$ points from the k-space image, $\T{DFT}{\times}\mathcal{I}$. Colorbar encodes matrix values, with white representing 0. Red rectangles in the DFT matrix indicate rows indexed by the random selection. \textit{Bottom panels:} Sampling matrix $\T{A}{=}\T{R}{\times}\T{DFT}$ illustrates the sampling operation including both random selection and Fourier transformation, connecting image $\mathcal{I}$ with $b$. }
\zfl{CS_equation_fig}
\end{figure}

The relation $\T{A} \mathcal{I}=b$ constrains $n$ variables with $m$ equations, thus the solution is under-determined. However we are interested in the solution that best satisfies: $\min _{\mathcal{I}}\left\{\| \mathcal{I}\|_{0}: \T{A} \mathcal{I}=b\right\}$. This refers to solving for an image $\mathcal{I}$ such that $\| \mathcal{I}\|_{0}$ is the smallest, subject to $\T{A} \mathcal{I}=b$, and where the $\ell_0$ norm is defined as the number of non-zero entries of a vector. However, $\ell_0$ minimization is a non-convex NP-hard problem in computation, therefore one usually solves its $\ell_1$ proxy.  An $\ell_1$ minimization produces an image with good proximity to the true image ($\mathcal{I} \approx \mathcal{\bar I}$), $\ell_1$ here serving as a relaxed version of $\ell_0$ minimization. We refer the reader to the fundamental papers~\cite{candes2006near, donoho2006compressed} for detailed explanation of the conditions under which a $\ell_1$ problem is equivalent to its $\ell_0$ counterpart.

In this case, the commonly employed constrained optimization problem can then be formally written as:
\begin{equation}
    \min _{\mathcal{I}}\left\{\| \mathcal{I}\|_{1}: \T{A} \mathcal{I}=b\right\}
\zl{CS_opt1}
\end{equation}
A typical image reconstruction process solves this $\ell_1$ minimization problem to find a solution image $\mathcal{I}$ (eg. \zfr{1d_support}B), although it may still differ slightly from the ``truth" $\mathcal{\bar I}$. In an optimization landscape, the solution $\mathcal{I}$ corresponds to the global $\ell_1$ minimum in the parameter space. When the global minimum (\zfr{1d_support}D) deviates from the actual (true) image (red in \zfr{1d_support}D), the reconstruction has reduced fidelity.

A standard procedure is to convert the constrained optimization to a standard linear programming (LP) problem in order to solve it. The $\ell_1$ norm in the problem is defined as: $\|\mathcal{I}\|_{1}=\sum_{i=1}^{n}\left|\mathcal{I}_{i}\right|$, where $\mathcal{I}_{i}$ represents the $i$th pixel of the image. For example, in \zfr{1d_support}A for instance, $\| \mathcal{I}\|_{1}$ is the sum of the absolute intensity of all 64 pixels in the 1$\times$64 image. Thus, the minimization finds the image $\mathcal{I}$ that yields smallest $\ell_1$ norm under the constraints of \zr{CS_opt1}. The computation process can be conducted by reformulating the problem into a LP one; for instance, the ``Basic Pursuit" (BP) problem in \zr{CS_opt1} is rewritten via the new set of parameters $u$ and $v$ as:
$$
\mathcal{I}=u-v, u_{i}=\max \left\{\mathcal{I}_{i}, 0\right\}, v_i=\max \left\{-\mathcal{I}_{i}, 0\right\}.
$$
The problem then becomes:
$$
\min _{u, v} \sum_{i=1}^{n} u_{i}+v_{i}
$$
subject to,
\begin{equation}
A(u-v)=b\: ;\: u \geq 0\:;\: v \geq 0.
\end{equation}
which is a linear programming problem.

From Candes and Tao et al.~\cite{candes2006robust}, we know that the number of measurements required to achieve high fidelity reconstruction is: $m = C\cdot s \text{log}(n)$; the corresponding probability is associated with $C$ in $1- \mathcal{O}(N^{-g (C)})$. This indicates that required $m$ to achieve a certain reconstruction probability largely depends on sparsity $s$. We adopt here the theorem below.

\T{Theorem (Tao 1)}~\cite{candes2006robust}: Assume a discrete signal $\mathcal{I}\in \boldsymbol{C}^{N}$ and a randomly chosen set of frequencies $\Omega$. Suppose $\mathcal{I}$ is a superposition of $|S|$ spikes ($\xd$-functions): $\mathcal{I} = \sum_{\tau\in S } \mathcal{I}(\tau)\delta (t-\tau)$ obeying:
\begin{equation}
|T| \leq C_{M} \cdot(\log N)^{-1} \cdot|\Omega|
\end{equation}
for some constant $C_{M}>0$. Then with probability at least $1- \mathcal{O}\left(N^{-M}\right)$, $\mathcal{I}$ can be reconstructed exactly as the solution to the $\ell_1$ minimization problem:
\begin{equation}
\min _{g} \sum_{t=0}^{N-1}|g(t)|, \text { s.t. } \hat{g}(\omega)=\mathcal{\hat{I}}(\omega) \text { for all } \omega \in \Omega
\end{equation}.
This theorem suggests to achieve reconstruction probability of $1- \mathcal{O}\left(N^{-M}\right)$, one needs at least $|S|\log (N)$ measurements. Here $|S|$ is the number of spikes, or equivalently the sparsity $s$.

\T{Theorem (Tao 2)} \cite{candes2006near}: 
Assume a vector $\mathcal{\bar I} \in \mathrm{R}^{N}$ that describes a digital signal or image. Suppose the $n$ th largest entry of vector $|\mathcal{\bar I}|$ obeys $|\mathcal{\bar I}|_{(n)} \leq R \cdot n^{-1 / p},$ where $R>0$ and $p>0$. Suppose one carries out measurements $b=\T{A}\mathcal{\bar I}$ where the $\T{A}_{k}$ ($k=1, \ldots, K$) are $N$ -dimensional Gaussian vectors with independent standard normal entries. Then for each $\mathcal{\bar I}$ obeying the decay estimate above for some $0<p<1$ and $K>(r\log N)^6$, our reconstruction $\mathcal{I}$ defined as the solution to the $\min _{\mathcal{I}}\left\{\| \mathcal{I}\|_{1}: \T{A} \mathcal{I}=b\right\}$, obeys
$$
\left\|\mathcal{I}-\mathcal{\bar I}\right\|_{\ell_{2}} \leq C_{p,\alpha} \cdot R \cdot(K / (\log N)^6)^{-r}, \quad r=1 / p-1 / 2
$$
with a probability at least $1-O\left(N^{-\rho / \alpha}\right)$. Here the constant $\rho$ is a function of $p$ and $\alpha$.

It is also convenient to introduce a quantity \textit{coherence} here to quantify the correlation between two different representations. If $(\Phi, \Psi)$ denotes a pair of orthonormal basis , then the coherence between $\Phi$ and $\Psi$ is defined to be,
$$
\mu(\Phi, \Psi)=\sqrt{n} \max _{1 \leqslant i, j \leqslant n}\left|\left\langle\phi_{i}, \psi_{j}\right\rangle\right|,
$$
where $\mu(\Phi, \Psi) \in[1, \sqrt{n} ]$. Suppose we have a vector that is represented in both spaces: $\mathcal{I} = \sum _{i}a_i \phi_i = \sum _{i}b_i \psi_i$. A small coherence suggests that an arbitrary coefficient $a_i$ for the vector in $\Phi$ space does not heavily rely on any single dimension in $\Psi$ space, given that the overlap between any of $\phi_{i}$ and $\psi_{j}$ is small. Instead, it contains information across all the dimensions from space $\Psi$, and vice versa. Indeed, when $\mu(\Phi, \Psi) = 1$, the two basis are said to are maximally incoherent.

In conventional CS employed in the context of MRI, $\Phi$ is the Fourier basis and $\Psi$ is the standard basis in x-space -- $[\Phi_1, ..., \Phi_n] = $\T{DFT}, and  $[\Psi_1, ..., \Psi_n] = \id$ (identity matrix). The two bases are therefore maximally incoherent. These ideas help in the generalization to dual-space CS approach as we describe below.

\subsection{Dual-space CS} 
We now provide more detail on the dual-CS methodology introduced in the main paper. Dual-space CS employs compressed measurements in both k- and real-space to produce higher fidelity images. In the dual-space CS program (as described in the main paper), a preliminary image is generated first through k-space measurements $b_k$ ($m_k\times$1) and $\ell_1$ minimization as outlined above. Subsequently, one samples real-space pixels with the highest intensity indicated by such preliminary image, and acquires signal $b_x$ ($m_x\times$1). We denote a subset in $b_x$ with non-zero entries to be $T$ ($|T|=t\leq m_x$). $T$ contains useful information for $\mathcal{I}$ because it consists of mostly non-zero entries. Combining information in both spaces, we employ a ``truncated optimization" strategy to reconstruct the image via dual-space CS. Truncation refers to removing the pixels obtained from real-space sampling so as to effectively increase the image sparsity from $s$ to $s{-}t$. Because the number of k-space measurements scales as $m= C\cdot s \text{log}(n)$, dual-CS is able to produce high fidelity image with the same number of measurements.

Consider for example, for the image considered in \zfr{1d_support}A with $n{=}$64, and $s {=}$6. Real-space sampling finds $t{=}$4 of them based on \zfr{1d_support}B, the number of unknowns reduces considerably from $s{=}$6 to $s{-}t{=}$2. This allows for a more rapid solution to the optimization problem, whose unknown dimension is largely reduced. Note that the exact number of samples conducted in k- and real-space $m_k$ and $m_x$ is carefully chosen to guarantee higher fidelity over single-space CS. 

As illustrated in \zfr{dual_equation_fig}, samples in both spaces build a ``truncated problem" where the set $T$ is removed from the optimization process. The image excluding the pixels that are indexed by $T$ is denoted as $\mathcal{I}_T$. We concatenate the measurements $b_k$ (\zfr{dual_equation_fig}A) and $b_x$ (\zfr{dual_equation_fig}B), to yield,
\begin{equation}
    \left[\begin{array}{l}b_x \\ b_k\end{array}\right] = \left[\begin{array}{l} \T{A}_x \\ \T{A}_k\end{array}\right] \times \mathcal{I},
\end{equation}
where $\T{A}_k$ and $\T{A}_x$ are a row-selected Fourier matrix and a row-selected identity matrix respectively. Since the $m_x$ variables are determined by $b_x$, we truncate them and solve other variables represented by $\mathcal{I}_{T}$. Truncation improves sparsity from $s{\rt}s-t$, and the data dimension of $\mathcal{I}_{T}$ is truncated to $n'{=}n{-}m_x$ (\zfr{dual_equation_fig}C). Accordingly $\T{A}'$ and $b'$ are then the truncated measurement matrix and measurement vector with dimension of $m_k\times (n-m_x)$ and $m_k \times 1$ (\zfr{dual_equation_fig}C). As a result, one can write the truncated problem as: 

\begin{equation}
    \min_{\mathcal{I}_T} \left\{\left\| \mathcal{I}_{T}\right\|_{1}: \T{A}' \mathcal{I}_T=b'\right\}.
\zl{trunc_opt1}
\end{equation}
This optimization is better constrained and easier to solve than \zr{CS_opt1} because the number of unknown variables can be considerably reduced (\zfr{1d_support}B2).

\begin{figure}[t]
  \centering
  {\includegraphics[width=0.5\textwidth]{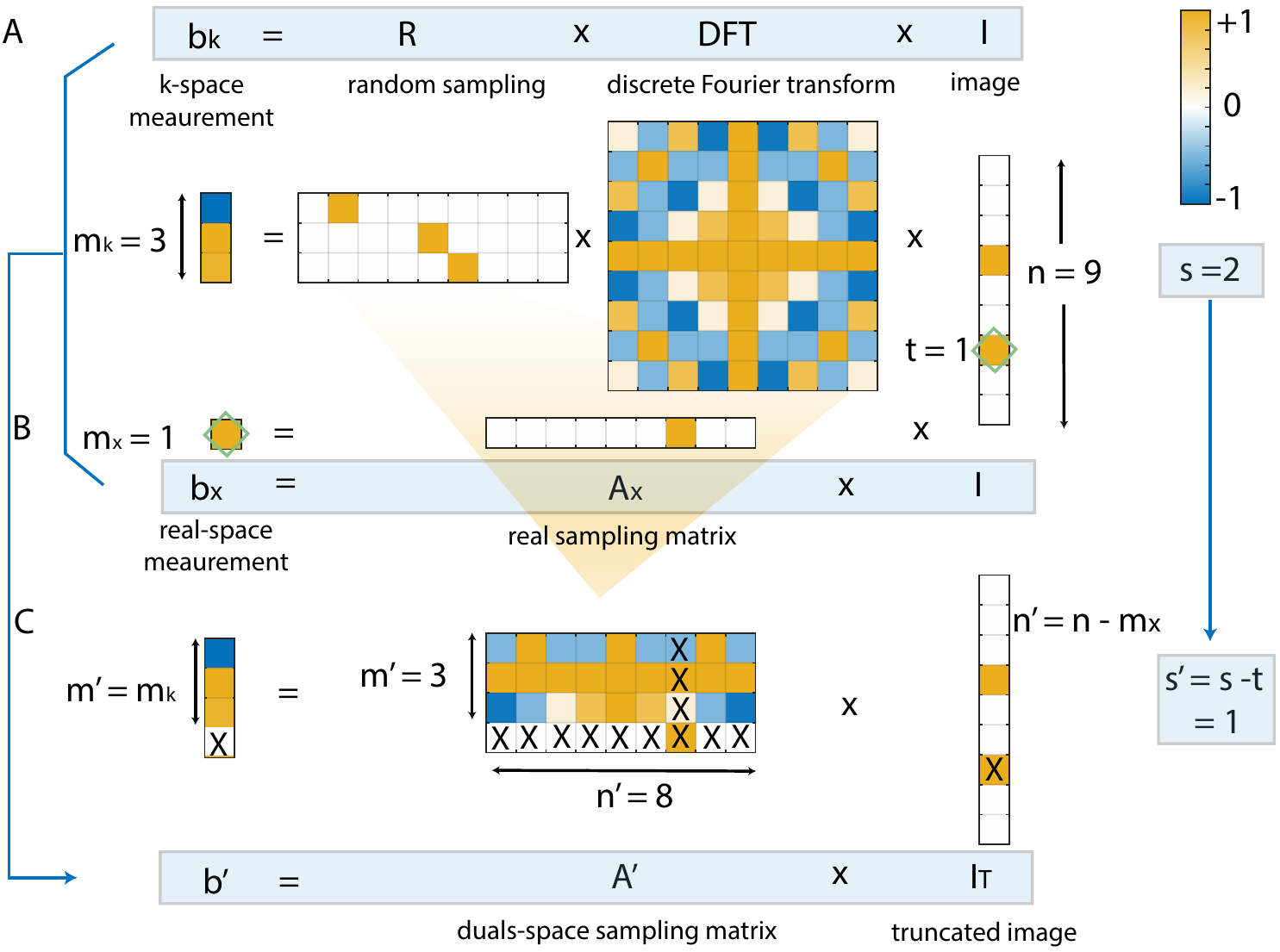}}
  \caption{\T{Schematic illustration of dual-space CS} in a matrix format, similar to \zfr{CS_equation_fig}. Here matrix values are represented graphically (see colorbar). In this example,  $m_k=3$ and $m_x=1$ measurements are carried out. Once again we assume a $n=9$ pixel image $\mathcal{I}$ with sparsity $s=2$. (A) \I{k-space sampling equation} similar to  conventional CS for $m_k$ measurements. (B) \I{Real-space sampling} equation representing a measurement taking place at the pixel of $i=7$ (green star). (C) \I{Dual-space sampling} equation now excludes the 7th dimension of the matrix (denoted here by crosses). The effective sparsity of the truncated image $\mathcal{I}_T$ is hence  enhanced from $s=2$ to $s_T=1$.}
\zfl{dual_equation_fig}
\end{figure}

\begin{figure}[t]
  \centering
  {\includegraphics[width=0.39\textwidth]{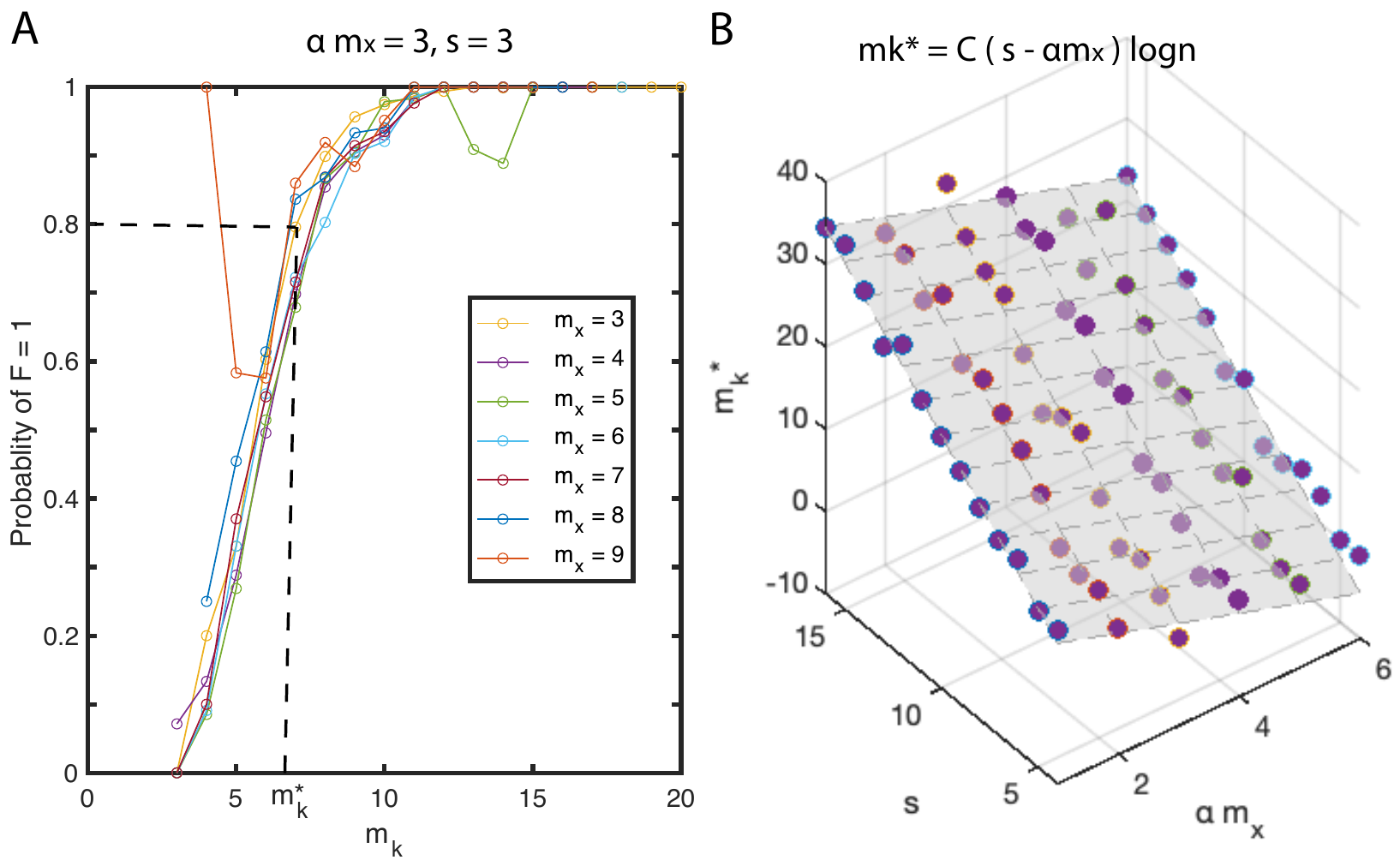}}
  \caption{\T{{Dependence of $m_k^*$ on $m_x$ and $s$. }} (A) \I{Convergence curves} for reconstruction with same $s-\alpha m_x$ but different $m_x$ overlap, indicating that required k-space measurement $m_k^*$ depends only on $s-\alpha m_x$. Here we define $m_k^*$ such that $\textrm{P}[\mathcal{F}=1] = 0.8$. The numerical results confirms that effective sparsity can be enhanced by $\alpha m_x$. (B) Numerics of $m_k^*$ as a function of $s$ and $\alpha m_x$ agree well with the plane of $m_k^* = C(s-\alpha m_x)\log n$. Each point here corresponds to 1000 times average.} 
\zfl{theory}
\end{figure}

\zfr{theory} describes numerics that confirm that the minimum number of measurements required for dual-space CS is $m^*_{\text{dual}} = C(s-\alpha_x m_x)\log n + m_x$. In these numerical experiments, images with a given sparsity $s$ are sampled with different $m_k$ and $m_x$; the success rate is then a function of $m_x$ and $m_k$, $\alpha_x = \alpha_x(m_x,m_k)$. Given that $m^*_{\text{dual}}$ depends on $\alpha_x$, it is apriori challenging to verify such relationship by simply varying $m_x$ without explicitly considering the functional dependence $\alpha_x(m_x,m_k)$. Instead here, we use a post-selection method when considering $\alpha_x m_x$ as a whole. Cases of $\alpha_x m_x (=s_T-s) = 3$ are post-selected, and plotted in \zfr{theory}A.  The plot illustrates that the probability of exact reconstruction ($\textrm{P}[\mathcal{F}=1]$) converges to 1 when $m_k$ increases. Curves with different $m_x$ overlap surprisingly well, indicating that $m_k^*$ depends solely on $s_T-s$. We select $\textrm{P}[\mathcal{F}=1] = 0.8$ to define $m_k^*$ (\zfr{theory}A). The choice of the required probability (0.8 here) only changes the prefactor $C$. The linear relationship between $m_k^*$ with both $s$ and $\alpha m_x$ in \zfr{theory}B also demonstrates that the effective sparsity is enhanced by $\alpha m_x$.

\section{Intuitive interpretation of dual-CS gains}
\label{intuitive}

To more intuitively understand the gains from the dual-CS approach, we offer three complementary viewpoints:
\benum
\item \I{Algorithmic perspective:} From the point of the view of the BP problem at the heart of the CS reconstruction, the proposed dual-space CS methodology solves a better constrained optimization problem. This is because the because during the truncation step after real-space sampling, the number of actual non-zero unknowns is reduced due to the excluding of detected non-zero entries. In these conditions, solving this truncated problem yields a higher probability for high fidelity image reconstruction.

\item \I{CS theory perspective:} Finding some non-zero entries in real-space lowers the effective sparsity by truncating the optimization problem. The effective image sparsity reduces by $t$ -- number of detected non-zeros. In conventional CS one requires $m = C\cdot s \text{log}(n)$ random measurements to obtain a good reconstruction. However, to now achieve a comparable reconstruction fidelity, one only needs $C\cdot (s-t) \text{log}(n-m_x) + m_x$ measurements in total, yielding a benefit. 

\item \I{Information perspective:} Simply considering this problem from an information point-of-view, for an image with high sparsity, the dimension of effective information is low, but the nominal dimension of the image data is high. One does not normally have access to the locations of the sparse entries in real-space; in conventional CS therefore a linear transformation of the information is collected in k-space. A more efficient approach is attempted in dual-space CS where the preliminary image obtained from conventional CS is employed as a guide to localize where the non-zeros are. Direct sampling them in real-space avoids searching in a large space, which is the primary motivation for CS techniques to be employed in the first place. 
\eenum

\begin{figure}[t]
  \centering
  {\includegraphics[width=0.4\textwidth]{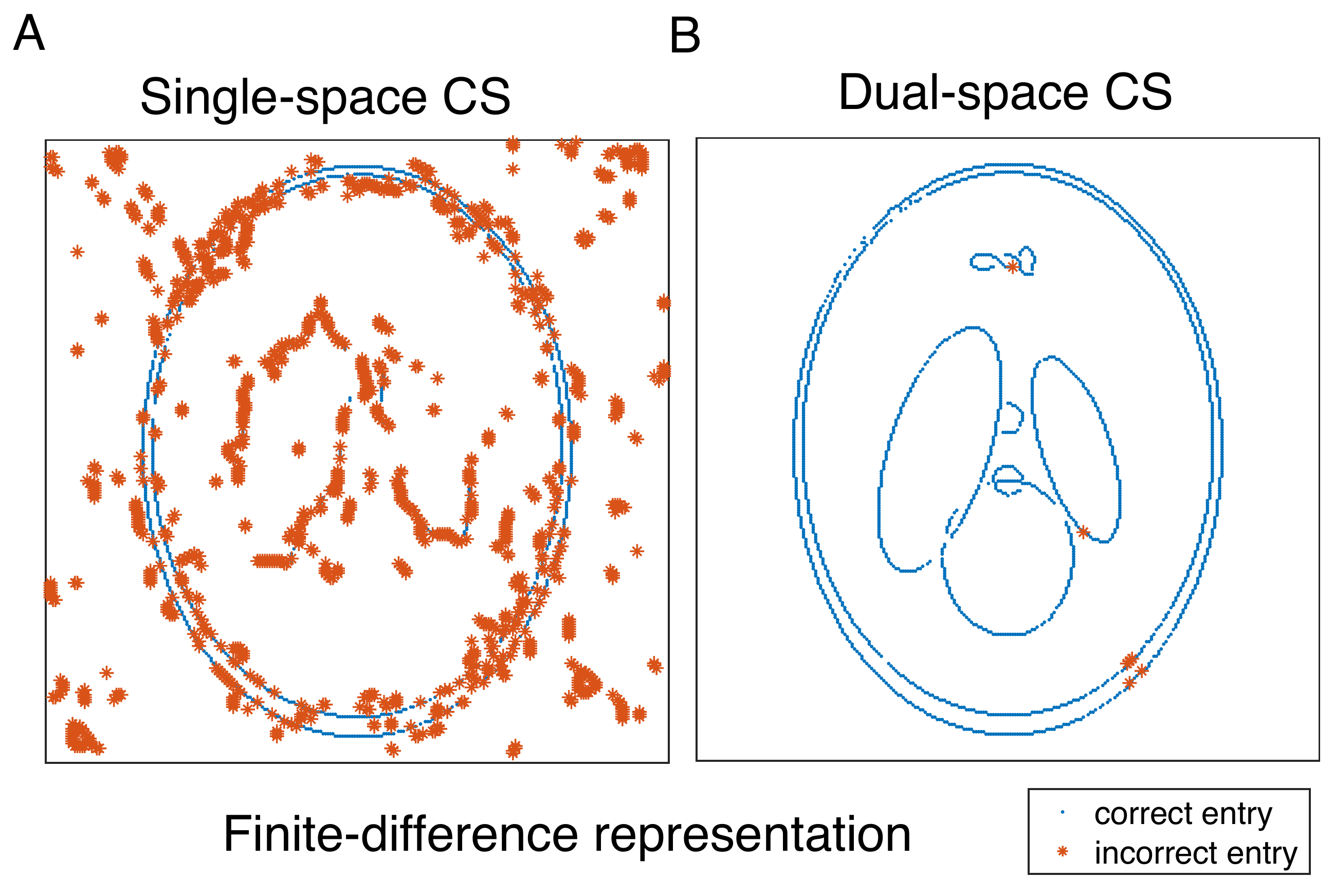}}
  \caption{\T{{Edge detection using finite-difference representation.}} (A-B) Shepp-Logan phantom (cf. \zfr{demo}) is displayed in the finite-difference representation after single-CS and dual-CS respectively. Blue dots represent true image edges while orange stars represent false positive edges.  Dual-space CS hence yields better edge reconstruction fidelity. Here $G_0{=}0.52$}
\zfl{fd_rep}
\end{figure}

\begin{figure*}[hbt!]
  \centering
  {\includegraphics[width=1\textwidth]{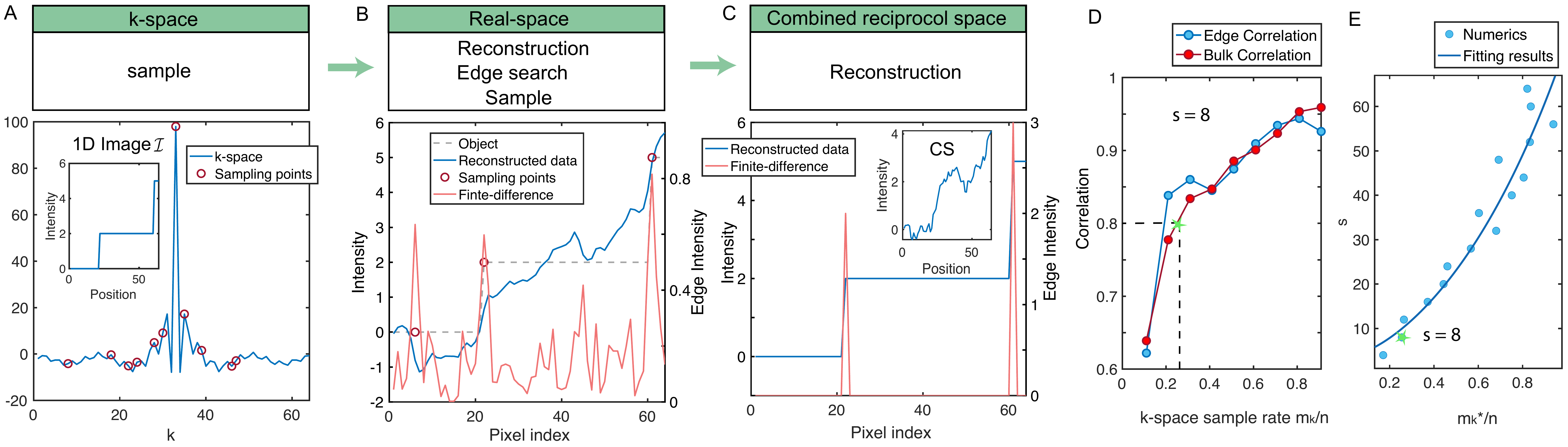}}
  \caption{\T{{Dual-space CS program for edge detection and figure of merit.}} Figure considers a representative image (shown in inset of (A)) with two edges to illustrate the dual-space CS program for edge detection. (A) \I{k-space random sampling:} $m_k =$ 11 red sample points are randomly sampled from the k-space image profile (blue). (B) \I{Real-space sampling:} The k-space data reconstructs a preliminary image (blue line) using conventional CS. In the finite-difference space (shown by red line), three points with largest gradients are identified as candidate edges for $m_x=$ 3 subsequent samples (red circles). (C) Combining information obtained from real- and k-space, the original image is reconstructed with high fidelity (blue line) and edges are located accurately. \I{Inset:} shows the result of image reconstruction with single-space CS with same number of total samples $m_x+m_k$. (D) \I{Edge and bulk correlation} with increasing number of samples $m_k$, shows that the reconstructed image rapidly converges to the true image with increasing k-space sample rate. Each point on the plot corresponds to an average over 50 random manifestation of 1-D images with sparsity $s=8$, and a random choice of sampling matrix.  \textit{Green star:} for clarity we denote a high-fidelity reconstruction as crossing a threshold correlation of $\mathcal{C}_e$ = 0.8. Vertical dashed line denotes corresponding value of $m_k^{*}/n$. (E) Panel shows the critical sample rate $m_k^{*}/n$ required for high fidelity reconstruction ($> 0.8$) for different values of sparsity $s$. These values are extracted from curves as in (D).    } 
\zfl{1d_illustration}
\end{figure*}


\section{Theoretical justification using t-NSP}
\label{sec_nsp}
In this section, inspired by Refs.\cite{zhang2005simple, zhang2005simple1}, we employ the Null Space Property (NSP) in order to formally demonstrate that dual-CS can aid in providing a high-fidelity reconstruction. In particular, we show that dual-CS provides a tighter bound represented by this property. An outline of the calculations presented in this section are as follows. First, we present NSP as a characteristic of the measurement matrix $\T{A}$ (Definition 1), following Ref. ~\cite{Cohen09}.  Following Ref.~\cite{Wang09},  we then formulate a variant of NSP for dual-CS (Definition 2) that we refer to as truncated null space property (t-NSP). In Theorem 1, we show that with certain images satisfying this property are guaranteed exact reconstruction. For a general image $\mI$, we then examine that under what conditions a random measurement $\T{A}$ will yield a high probability reconstruction (Theorem 2).  The proof here is reproduced from Ref. ~\cite{Wang09} for completeness. Finally, we compare the situation before and after real-space measurement and show that the reconstruction probability is improved by dual-space CS (Theorem 3). This NSP scheme is also useful in analyzing the scaling of the dual-CS program. 

For completeness, we reproduce Definition 1 and Definition 2 below from Ref.\cite{Cohen09} and Ref.\cite{Wang09}.

\noindent\textit{Definition 1: Null Space Property} -- \\
A matrix \(A \in \mathbb{R}^{m \times n}\)  satisfies the NSP of order $L$ for $\gamma >0$ if 
\begin{equation}
\left\|\eta_{S}\right\|_{1} \leq \gamma\left\|\eta_{S^C} \right\|_{1}
\end{equation}
holds for all index sets $S$ if \(|S| \leq L\) and all \(\eta \in \mathcal{N}(\T{A})\), the null space of $\T{A}$. Here $S^C$ is a complement of S with respect to the full set $\{1, . . . , n\}$.

\noindent\textit{Definition 2: Truncated Null Space Property} -- \\
A matrix \(A \in \mathbb{R}^{m \times n}\)  satisfies the t-NSP of order $L$ for $\gamma >0$ and $0< t  \leq n$ if 
\begin{equation}
\left\|\eta_{S}\right\|_{1} \leq \gamma\left\|\eta_{\left(T \cap S^{C}\right)}\right\|_{1}
\end{equation}
holds for all index sets \(T \subset\{1, \ldots, n\}\) with \(|T|=t\), all subset $S\subset T$ with \(|S| \leq L\) and all \(\eta \in \mathcal{N}(\T{A})\), the null space of $\T{A}$.
For simplicity, we use t-NSP$(t, L, \gamma)$ to denote the t-NSP of order $L$ for $\gamma$ and $t$, and use $\gamma$ to replace $\gamma$ and write t-NSP$(t, L, \gamma)$ where $\gamma$ is the infimum of all the feasible $\gamma$ satisfying the above condition.

Similar to a more well known property -- Restricted Isometry Property (RIP)~\cite{candes2006robust}, NSP condition for the sampling matrix is used to determine wether there is a unique solution for the reconstruction problem. To give readers more intuitive understanding, we can specify the variables above in the dual-CS setting. First, the null space is important in that: if we know a solution $\T{A}x_0{=}b$, then a general solution to $\T{A}\mI{=}b$ is $\mI{=}x_0{+}\eta, \eta \in \mathcal{N}(\T{A})$. Second, $\eta_S$ can be considered as a non-zero vector in the sparse representation, so that $S$ is an index subset of all the non-zero entries. Definition 1 suggests, given any image (with a specific sparse set $S$), the $\ell{-}1$ norm of these non-zero entries will not concentrate on $L$ elements. The level of distribution is quantified by $\gamma$. Third, $\eta_T$ will be a  entry after truncation and $T$ is the index set of the truncated space dimensions. t-NSP is similar to NSP, but in a truncated space.
\\

\noindent\textit{Theorem 1: Reconstruction} --\\
Let \(\bar{\mI}\) be a given vector that satisfies $\T{A}\bar{\mI}{=}b$, \(\bar{S}=\left\{i: \bar{\mI}_{i} \neq 0\right\},\) and \(T\) be given such that \(T \cap \bar{S} \neq \emptyset .\) Assume that a matrix \(\T{A}\) satisfies t-NSP\((t, L, \gamma)\) for \(t=|T| .\) If \(\left\|\bar{\mI}_{T}\right\|_{0} \leq L\) and \(\gamma<1,\) then \(\bar{\mI}\) is the unique minimizer of optimization problem \zr{CS_opt1}.

\noindent Proof:
The proof is inspired by Ref.\cite{zhang2005simple, zhang2005simple1,Wang09}. Define $ S:=T \cap \bar{S}$ .  We take $\forall \eta \in \mathcal{N}(A)$.
\begin{equation}
\begin{aligned}  
\left\|\bar{\mI}_{T}+\eta_{T}\right\|_{1} &=\|\bar{\mI}_{S} +  \bar{\mI}_{T \cap S^{C}} + \eta_{S}+\eta_{T \cap S^{C}}\|_1 \\
&=\left\|\bar{\mI}_{S}+\eta_{S}\right\|_{1}+\left\|\mathbf{0}+\eta_{T \cap S^{C}}\right\|_{1} \\ 
\end{aligned}
\end{equation}

Since $\left\|\bar{\mI}_{S}\right\|_{1}=\left\|\bar{\mI}_{T}\right\|_{1}$, we can re-write the above:
\begin{equation}
\begin{aligned}  
\left\|\bar{\mI}_{T}+\eta_{T}\right\|_{1} &= \left\|\bar{\mI}_{T}\right\|_{1}+ \underbrace{\left(\left\|\bar{\mI}_{S}+\eta_{S}\right\|_{1}-\left\|\bar{\mI}_{S}\right\|_{1}+\left\|\eta_{S}\right\|_{1}\right)}_{\geq 0}\\
&+\left(\left\|\eta_{T \cap S^{C}}\right\|_{1}-\left\|\eta_{S}\right\|_{1}\right)
\end{aligned}
\zl{v1}
\end{equation}

For the last two terms in the above, we have \(\left\|\eta_{T \cap S^{C}}\right\|_{1}-\left\|\eta_{S}\right\|_{1}>0\), since \(\left\|\eta_{S}\right\|_{1} <\gamma\left\|\eta_{T \cap S^{C}}\right\|_{1}<\left\|\eta_{T \cap S^{C}}\right\|_{1}\) given the definition of t-NSP$(|T|, L, \gamma)$.

$\left\|\bar{\mI}_{T}\right\|$ is the unique solution for the optimization problem, if and only if:
\begin{equation}
\left\|\bar{\mI}_{T}+\eta_{T}\right\|_{1}>\left\|\bar{\mI}_{T}\right\|_{1}, \forall \eta \in \mathcal{N}(A), \eta \neq 0
\zl{v}
\end{equation}
And we have already proved \zr{v} in \zr{v1}.

Intuitively, considering the setting of dual-CS, $\mI$ is a solution of dual-CS, and $\bar{S}$ indexes its non-zero entries, while $S$ corresponds to its index set after truncation. Thus, $\left\|\bar{\mI}_{S}\right\|_{1}=\left\|\bar{\mI}_{T}\right\|_{1}$. Theorem 1 suggests that t-NSP and $\gamma < 1$ leads to an unique exact reconstruction.
\\

\noindent\textit{Theorem 2: High probability reconstruction} --\\
If matrix $\T{A} \in \mathbb{R}^{m \times n}$ satisfy t-NSP$(t, L, \gamma)$ and $\gamma < 1$, then with probability higher than certain value, $\mathcal{\bar I}$ is the unique solution to the problem in \zr{trunc_opt1}, if $s<\mathcal{S}(m_k, m_x, n)$, where $\mathcal{S}(m_k, m_x, n) = C_0 \frac{m_k-m_x}{1+\textrm{log}(\frac{n-m_x}{m_k-m_x})}+m_x$.

This is a variation of a theorem presented in Ref.\cite{Wang09}. It implies that smaller $\gamma$ corresponds to a tighter condition in the t-NSP and when $\gamma{<}1$, we obtain a bounded high-fidelity reconstruction.\\

\noindent\textit{Theorem 3: $\gamma$-reduction} --\\
If $\T{A} \in \mathbb{R}^{m \times n}$ satisfy t-NSP$(t, L, \gamma)$ as well as t-NSP$(t', L', \gamma')$ with $t'<t$ and $L'<L$. If $(L-L')>\gamma(t=t'-(L-L'))$, then $\gamma' < \gamma$.

For the complete proof we refer the reader to Ref.\cite{Wang09}. This theorem suggests that the reduction of $\gamma$ in dual-CS in comparison to single-CS leads to a tighter t-NSP condition, and better reconstruction.

\section{Edge detection}
\label{sec_edge}
\subsection{Algorithm for dual-CS edge detection}
In this section, consider in greater detail the edge detection algorithm employing dual-CS as presented in the main paper. \zfr{fd_rep} gives an example of such an edge-detection implementation exploiting the fact that the image is highly sparse in FD space. Single CS is inefficient in sampling entries in this space, yielding a a low fidelity reconstruction (\zfr{fd_rep}A) (highlighted by the orange points). Dual-CS yields performs much better in comparison. 

Let us now consider the implementation of dual-CS edge detection algorithm in more detail.  The convex optimization problem (\zr{CS_opt1}) and problem (\zr{trunc_opt1}) are solved in step (\T{\rom{2}}) and (\T{\rom{4}}) respectively. Step (\T{\rom{2}}) is similar to conventional CS except that fewer measurements are taken. In step (\rom{4}), we define a weighted total variation function (wTV) of a image $u$ to be 
\begin{equation}
\text{wTV}(\mI)=\sum_{\alpha} g_{\alpha}\left|D_{\alpha} \mI \right|:=\sum g_{(i, j) \sim(k, l)}\left| \mI_{i, j}-\mI_{k, l}\right|
\label{wTV}
\end{equation}
Here the sum is taken over all pixel pairs $(i, j)$ and $(k, l)$, and the weight $g_{(i, j) \sim(k, l)}$ is either 1 or 0 representing non-edge and edge respectively. This function is initialized with $g_{(i, j) \sim(k, l)} = 1$ for all the pairs, and updates to 0 when real space sampling detects a edge between a specific pair $(i, j)$ and $(k, l)$. Considering $\sum_{\alpha} g_{\alpha}\left|D_{\alpha} \mI \right|$ as a whole to be $\mI_T$, minimizing such Eq. (\ref{wTV}) becomes a truncated $\ell_{1}$ \text minimization problem in \zr{trunc_opt1}. The minimization problem subject to equality constraint can be written in a Lagrangian form and solved protocol step (\T{\rom{4}}).

We illustrate the dual-CS edge detection program with a simple 1-D example and demonstrate fidelity gains over single space CS. An arbitrary 1-D image as a linear combination of multiple Heaviside step functions: $\mathcal{I}(x) = \sum_{i} \alpha_i \textrm{sgn} (x-x_i)$, where $\{x_i\}$s are the step positions. We consider a simple case of a 2-step signal (\zfr{1d_illustration}A inset), and its k-space representation is shown in \zfr{1d_illustration}A.

\zfr{1d_illustration} describes the k- and real-space measurements, each followed by a reconstruction step, jointly producing high fidelity signals. The initial compressed sampling ($m_k$ measurements represented in \zfr{1d_illustration}A) of the k-space and reconstruction (blue line in \zfr{1d_illustration}B) resembles a typical CS in MRI~\cite{lustig2007sparse}. The reconstructed signal preserves the bulk shape of the object (dashed line \zfr{1d_illustration}B), but without high fidelity. Especially the edges are not clearly resolved. We feed such preliminary image to guiding subsequent sampling in real-space; $m_x$ candidate ``edge" locations (red circles \zfr{1d_illustration}B) are calculated from the preliminary image and inspected by real-space measurements.

Combining data from two spaces by the truncated optimization, we reconstructed an image (\zfr{1d_illustration}C) completely identical to the original. Here the truncation removes the detected edges from the optimization problem given that they are no longer knowns. If one instead employs same number of samples $m_k+m_x$ solely in k-space, the CS protocol generates a lower fidelity image in \zfr{1d_illustration}C inset.

\subsection{Performance analysis}
\label{sec_performance}
\begin{figure}[t]
  \centering
  {\includegraphics[width=0.46\textwidth]{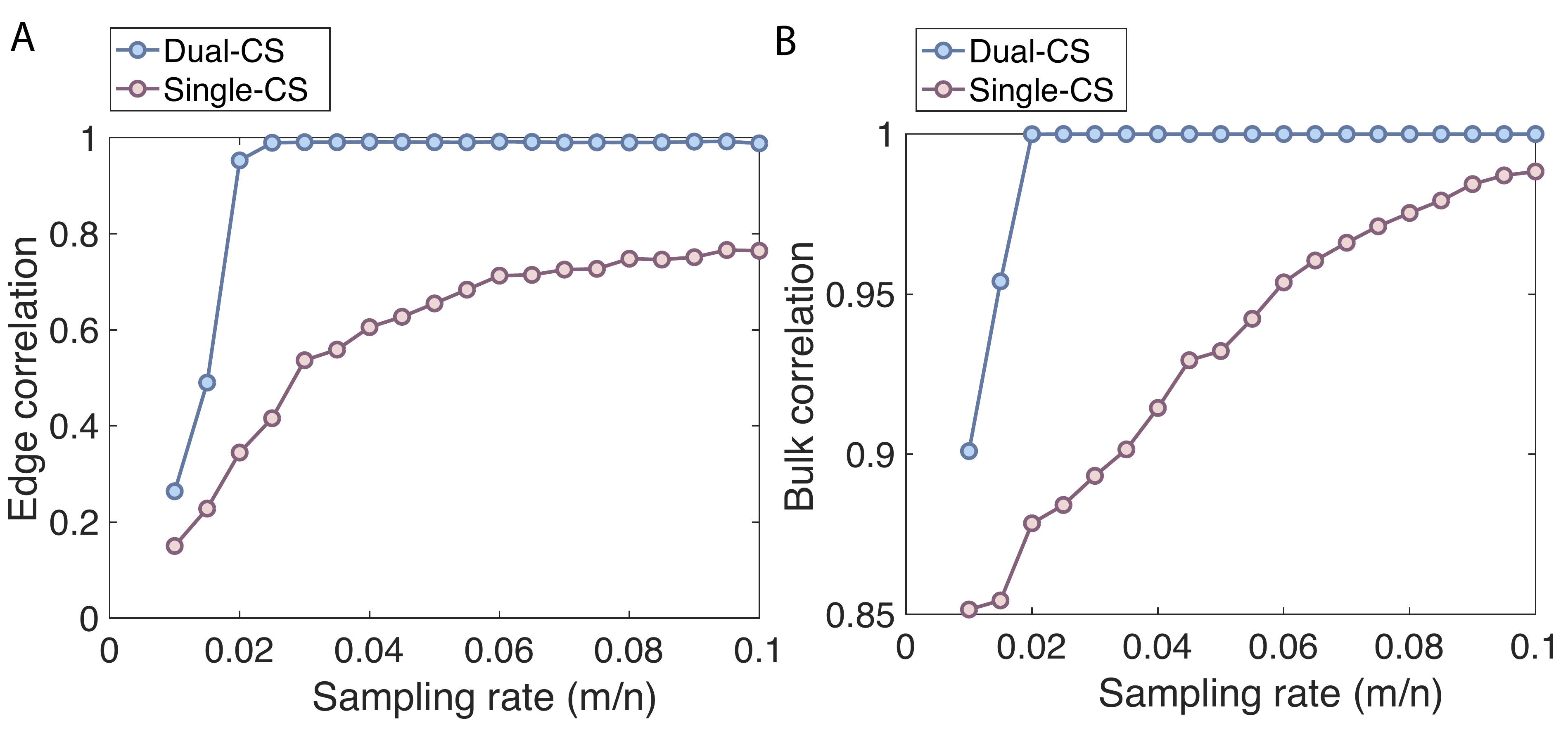}}
  \caption{\T{Quantifying performance of dual-space CS.} Panels show the edge and bulk correlation for the 256 $\times$ 256 Shepp-Logan phantom in Fig. 4 of the main paper as a function of sampling rate $m/n$ for dual-space CS (blue) and single-CS (purple). (A) \I{Convergence of edge image correlation} shows that dual-space CS promotes faster convergence. (B) \I{Convergence of bulk image correlation} is similarly improved  by using the information of edges and k-space sub-sampled to promote faster convergence in the dual-space CS case (blue curve). }
\zfl{compare}
\end{figure}

\begin{figure}[t]
  \centering
  {\includegraphics[width=0.46\textwidth]{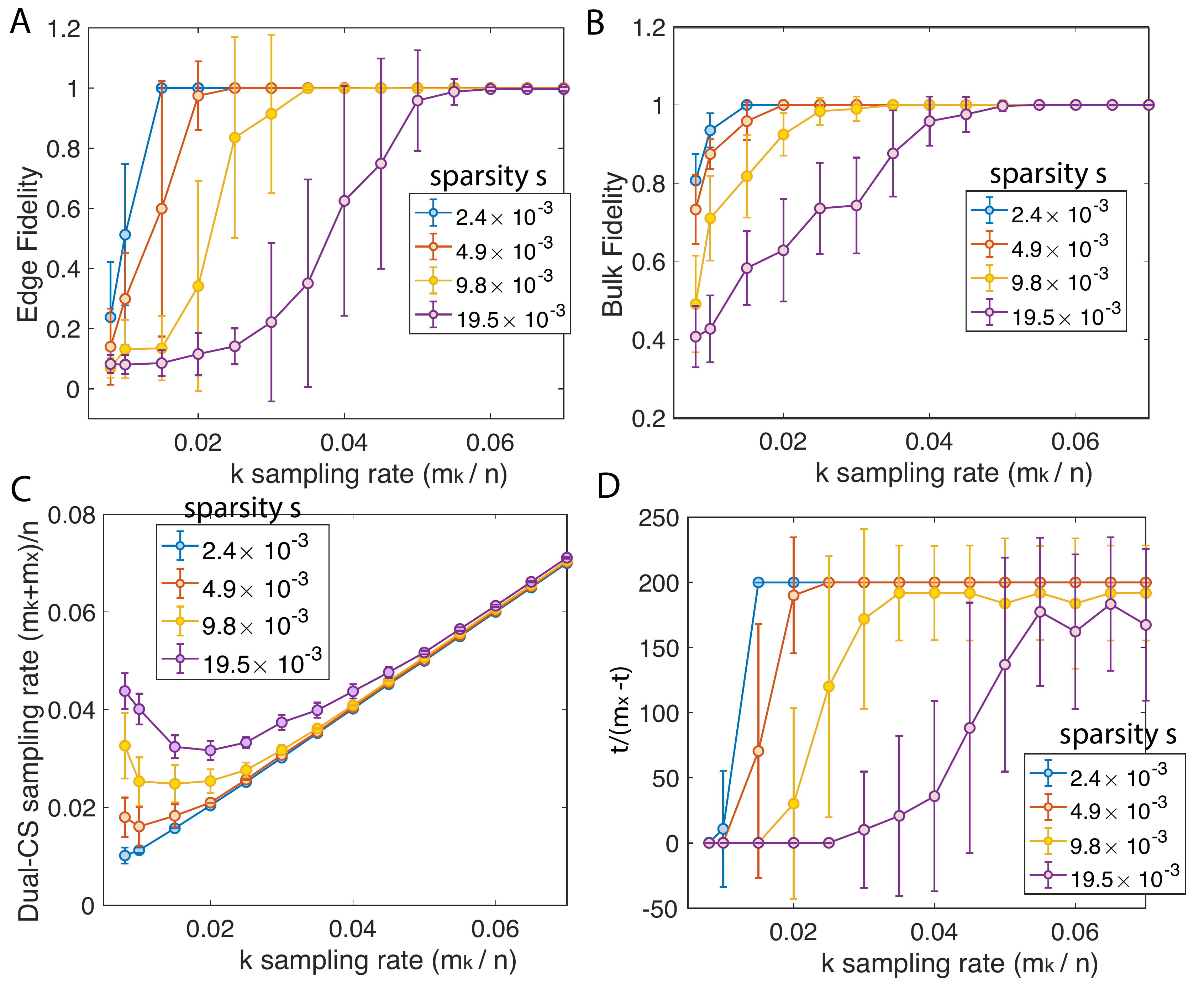}}
 \caption{\T{Dual-space CS reconstruction for images of different sparsity} Synthesized images of 64$\times$64 are employed here for numerics, and each point in the panels corresponds to averaging 30 manifestations of the randomly generated images. (A) \I{Edge detection performance} for images of with different sparsity are plotted against k-space sampling rate. Turning point of each curve scales with sparsity in an approximately square-root fashion. (B) \I{Bulk image reconstruction performance} plotted for the images of different sparsity. (C) \I{Total sampling rate} required for a high-fidelity reconstruction. When k-space sampling rate is low, one would require higher real-space sampling to ensure best performance. (D) Ratio between true positive detection $t$ and false positive detection $m_x{-}t$.}
\zfl{analysis}
\end{figure}

Varying sampling ratio in k-space, we find a significantly faster convergence of reconstruction fidelity for our protocol when compared with conventional CS, for both bulk (\zfr{compare}A) and edge (\zfr{compare}B) components. Here we characterize image fidelity through correlation between the reconstruction and the object: $\mathcal{C}=\sum(\mathcal{I}-\langle\mathcal{I}\rangle)\left(\mathcal{\bar I}-\left\langle \mathcal{\bar I}\right\rangle\right)$, where $\langle\cdot\rangle$ indicates the mean value. The edge and bulk correlation $\mathcal{C}_e$ and $\mathcal{C}_b$ can be defined when substituting $\mathcal{I}_e$ or $\mathcal{I}_b$ in the definition. 

To understand the fidelity gains in more general cases than this example, we illustrate the relationship between image fidelity and sample number $m_k$, $m_x$. As elucidated in Sec.~\ref{sec_nsp}, a high probability reconstruction requires measurement number to satisfy the inequality: $s<\mathcal{S}(m_k, m_x, n)$, where $s$ is image sparsity. The function, $\mathcal{S}(m_k, m_x, n) = c_0 \frac{m_k-m_x}{1+\textrm{log}(\frac{n-m_x}{m_k-m_x})}+ \alpha_x m_x$ ($c_0$ constant), quantifies the reconstruction probability. When $m_x=$0, the condition reduces to conventional CS: $s< c_0\frac{m_k}{1+\textrm{log}(n/m_k)}$~\cite{zhang2013theory}. Given a fixed total sample number $m = m_k + m_x$, a careful balance of $m_k$ and $m_x$ allows for larger $\mathcal{S}$ compared with the case of $m_x {=} 0$. Thus, in the relationship between image fidelity and sample number above, image fidelity peaks at $m_x \neq 0$; the best sampling scheme contains both k- and real-space data.
We numerically verify this relationship laid out above by examining the fidelity phase transition. As shown in \zfr{1d_illustration}D, given sparsity $s =$ 8, both bulk and edge fidelity increase as the k-space sample $m_k$ advances, and asymptotically converge after a crossover point. We extract this crossing point $m_k$, where the condition $s = \mathcal{S}$ is satisfied. Simulating multiple convergence curves with different $s$, critical $m_k$ are plotted on \zfr{1d_illustration}E. The fitting curve $s= c_0 \frac{m_k-m_x}{1+\textrm{log}(\frac{n-m_x}{m_k-m_x})}+ \alpha_x m_x$, where $m_x =$ 5 and $n = $ 64 agrees well with numerics, supporting the statement above. Notably, there is only one free parameter $c_0$ throughout the fitting.

Note here, it is challenging to derive the required $m_x$ and $m_k$ for certain $s$ and $n$ given that $\alpha_x$ strongly depends on the structural feature of the images. However, the above $\mathcal{S}$ function provides the scaling of the maximum sparsity of an image with respect to $m_x$, $m_k$ and $n$.
We further analyzed the strategy performance when varying image sparsity and number of k/real-space sampling (\zfr{analysis}). Higher $s$ number implies more complicated images, and thus requires more number of samples to converge (\zfr{analysis}A, B). \zfr{analysis}C shows the total sampling rate required for high fidelity reconstruction, and \zfr{analysis}D quantifies the extent of false positive detection with k-space sampling rate.

\end{document}

%% file: Commands3.tex
\newcommand{\id}{\mathds{1}}

\newcommand{\xd}{\delta}

\newcommand{\app}{\approx}

\newcommand{\Cs}{{}^{13}\R{C}}

\newcommand{\mI}[0]{\mathcal{I}}

\newcommand{\mO}[0]{\mathcal{O}}

\newcommand{\ov}[1]{\overline{#1}}

\newcommand{\sq}[1]{\sqrt{#1}}

\newcommand{\mF}[0]{\mathcal{F}}

\newcommand{\rt}{\rightarrow}

\newcommand{\beq}{\begin{equation}}
\newcommand{\eeq}{\end{equation}}
                  
\newcommand{\benum}{\begin{enumerate}}
\newcommand{\eenum}{\end{enumerate}}
                    
\newcommand{\bit}{\begin{itemize}}
\newcommand{\eit}{\end{itemize}}

\newcommand{\bea}{\begin{eqnarray}}
\newcommand{\eea}{\end{eqnarray}}

\newcommand{\bev}{\begin{verbatim}}
\newcommand{\eev}{\end{verbatim}}

\newcommand{\zt}{\times}

\newcommand{\T}[1]{\textbf{#1}}
\newcommand{\I}[1]{\textit{#1}}
\newcommand{\R}[1]{\textrm{#1}}

\newcommand{\zl}[1]{\label{eqn:#1}}
\newcommand{\zr}[1]{Eq.\,(\ref{eqn:#1})}
\newcommand{\zfl}[1]{\protect\label{fig:#1}}
\newcommand{\zfr}[1]{\figurename\,\ref{fig:#1}}

\newcommand{\braket}[2]{\langle{#1}\vert{#2}\rangle}

\newcommand{\ba}{\left\{ \begin{array}{lr}}
\newcommand{\ea}{\end{array}\right.}

\newcommand{\blist}[1]{
 \begin{list}{#1}
 \begin{align}
	 arrow
 \end{align}
 $\checkmark\star
  { \setlength{\itemsep}{3pt}
     \setlength{\parsep}{2pt}
     \setlength{\topsep}{3pt}
     \setlength{\partopsep}{0pt}
     \setlength{\leftmargin}{1em}
     \setlength{\labelwidth}{1em}
     \setlength{\labelsep}{0.5em} } }
\newcommand{\elist}{
  \end{list}  }

\DeclareMathSymbol{\vartheta}{\mathalpha}{letters}{"12}
\DeclareMathSymbol{\theta}{\mathalpha}{letters}{"23}
\DeclareMathSymbol{\phi}{\mathalpha}{letters}{"27}
\DeclareMathSymbol{\varphi}{\mathalpha}{letters}{"1E}

\newcommand{\bef}
{
\begin{figure}[htbp]
\centering
}

\newcommand{\eef}{\end{figure}}
